\documentclass[aps,prb,twocolumn,showpacs,notitlepage,floatfix,longbibliography,superscriptaddress]{revtex4-1}

\usepackage{amsmath}
\usepackage{amsthm}
\usepackage{amssymb}
\usepackage{mathtools}
\usepackage[all]{xy}
\input xy
\xyoption{matrix}
\xyoption{2cell}
\xyoption{arrow}
\xyoption{curve}
\xyoption{all}
\usepackage{graphicx}
\usepackage[colorlinks=true,linkcolor=blue,citecolor=blue,filecolor=blue,urlcolor=blue,pdfstartview=FitH,breaklinks=true]{hyperref}
\usepackage{xcolor}
\usepackage[normalem]{ulem}

\newcommand{\pksadd}{Max Planck Institute for the Physics of Complex Systems, Dresden D-01187, Germany}
\newcommand{\pcsadd}{Center for Theoretical Physics of Complex Systems, Institute for Basic Science(IBS), Daejeon 34126, Korea}
\newcommand{\ustadd}{Basic Science Program(IBS School), Korea University of Science and Technology(UST), Daejeon 34113, Korea}
\newcommand{\umassadd}{Department of Mathematics and Statistics, University of Massachusetts, Amherst MA 01003-4515, USA}

\newcommand{\mh}{\mathcal{H}}
\newcommand{\hmh}{\hat{\mh}}

\begin{document}

\title{Nonlinear caging in All-Bands-Flat Lattices}

\author{Carlo Danieli}
\affiliation{\pksadd}
\affiliation{\pcsadd}

\author{Alexei Andreanov}
\affiliation{\pcsadd}
\affiliation{\ustadd}

\author{Thudiyangal Mithun}
\affiliation{\pcsadd}
\affiliation{\umassadd}

\author{Sergej Flach}
\affiliation{\pcsadd}
\affiliation{\ustadd}

\date{\today}

\begin{abstract}
    We study the impact of classical short-range nonlinear interactions on transport in lattices with no dispersion. The single particle band structure of these lattices contains flat bands only, and cages non-interacting particles into compact localized eigenstates. We demonstrate that there always exist local unitary transformations that detangle such lattices into decoupled sites in dimension one. Starting from a detangled representation, inverting the detangling into entangling unitary transformations and extending to higher lattice dimensions, we arrive at an All-Bands-Flat generator for single particle states in any lattice dimension. The entangling unitary transformations are parametrized by sets of angles. For a given member of the set of all-bands-flat, additional short-range nonlinear interactions destroy caging in general, and induce transport. However, fine-tuned subsets of the unitary transformations allow to completely restore caging. We derive the necessary and sufficient fine-tuning conditions for nonlinear caging, and provide computational evidence of our conclusions for one-dimensional systems.

\end{abstract}

\maketitle

\section{Introduction}

Understanding the impact of interactions on single particle localized states has been one of the most intriguing quests of the past decades in condensed matter physics. Classical and quantum approaches may yield seemingly distinct outcomes while starting from the same single particle localization. One notable example concerns the impact of interactions on Anderson localization -- i.e. the exponential localization of all single particle states due to uncorrelated disorder and the confinement of noninteracting particles over finite portions of the lattice.~\cite{anderson1958absence,kramer1993localization} Weakly interacting quantum particles show a finite temperature transition from a thermalized to a many-body localized phase.~\cite{basko2006metal,aleiner2010finite,abanin2019colloquium} Classical interactions instead predict finite heat and particle conductivity at arbitrarily small temperatures, and related indefinite subdiffusive wave-packet spreading.~\cite{laptyeva2014nonlinear,vakulchyk2019wave}

In translationally invariant networks destructive interference can fully localize subfamilies of single particle eigenstates within a finite portion of the lattice. These eigenstates -- dubbed \emph{compact localized states} (CLS) -- have macroscopically degenerate eigenenergies and form disperionless (or flat) Bloch bands in band structures containing otherwise dispersive bands. While they were originally used to study degenerate ferromagnetic ground states,~\cite{mielke1991ferromagnetic,tasaki1992ferromagnetism} \emph{flatband lattices} have drawn a lot of theoretical attention ever since and compact localized states have been observed experimentally in several settings, from ultracold atomic gases to photonics -- for an overview on recent advances see Refs.~\onlinecite{derzhko2015strongly,leykam2018artificial,leykam2018perspective}. Flatband networks can be viewed as fine-tuned submanifolds in a suitably defined space of Hamiltonian tight-binding networks.

Remarkably, flatband networks can be further fine-tuned in order to flatten remaining dispersive bands, all the way down to 
an important subclass of the family of flatband lattices which possess only perfectly flat Bloch bands. 
The absence of single particle dispersion yields the strict confinement of noninteracting particles within the lattice. This was first shown in a two-dimensional lattice structure in presence of a magnetic field which was fine-tuned to reach a time-reversal invariant model,~\cite{vidal1998aharonov} and was dubbed \emph{Aharonov-Bohm} caging (AB), while the compact eigenstates were referred to as \emph{caged states}. This path to reach single particle caging has been extended further in the past 
decade~\cite{doucot2002pairing,fang2012photonic,longhi2014aharonov,kibis2015aharonov,hasan2016optically} and it has been experimentally realized using photonic lattices~\cite{fang2012realizing,mukherjee2018experimental} and qubits nano-circuits,~\cite{gladchenko2009superconducting} among others. Interestingly the introduction of a magnetic field, which in general does break time reversal, is not of essence and not needed at all, as we will show below. The route to zero dispersion and caging via magnetic fields leads to one model realization among whole manifolds of systems which lack dispersion. We coin such systems \emph{all bands flat} (ABF) lattices.

The impact of interactions on the ABF single particle caging has been studied in a number of attempts in both classical and quantum regimes. A notable set-up for these studies has been the 1D diamond (rhombic) ABF chain. In this case, while local Kerr nonlinear interactions preserve caging,~\cite{gligoric2018nonlinear,diliberto2019nonlinear} Hubbard interactions induce transporting bound states of two particles simultaneously.~\cite{vidal2000interaction}
The study of quantum interactions for caged noninteracting particles has been further developed in different ABF geometries, including the 1D Creutz lattice~\cite{creutz1999end,takayoshi2013phase,tovmasyan2013geometry,tovmasyan2016effective,junemann2017exploring,tovmasyan2018preformed} and the 2D Dice lattice.~\cite{vidal2001disorder} 
 
In this work, we study classical nonlinear interactions in ABF networks. We will obtain conditions under which an additional fine-tuning in the manifold of ABF networks leads to a complete caging in the presence of nonlinear interactions.  We will show that in the absence of interactions, proper local unitary transformations lead to a complete detangling of the network into decoupled sites for 1D systems -- a fact which holds for any number of bands and provides a systematic (in any lattice dimension) and exhaustive (at least in 1D) generator for ABF lattices. We employ these unitary transformations to show that nonlinear interactions in general break the single particle caging and result in transport and delocalization. We then obtain necessary and sufficient fine-tuning conditions for the nonlinear interaction to preserve caging. The condition is then tested for 1D networks with $\nu=2,3,4$ flatbands, which will include previously studied diamond chain ($\nu=3$) examples, for which the nonlinear caging was previously found.~\cite{gligoric2018nonlinear,diliberto2019nonlinear} We further present extensions to 2D nonlinear caging models. The intricate related caging features arising from quantum interactions will be unfolded in a subsequent work.~\cite{danieli2020quantum}

\section{Single particle caging}
\label{sec:abf-gen}

Let us consider the unitary evolution of a one-dimensional tight-binding problem with nearest neighbour unit cell coupling:
\begin{gather}
    i\dot{\psi}_n = - H_0\psi_n -  H_1\psi_{n+1} - H_1^\dagger\psi_{n-1}.
    \label{eq:fb_eq}
\end{gather}
For any $n\in\mathbb{Z}$, each component of the complex vector $\psi_n=(\psi_{n,1},\dots,\psi_{n,\nu})^T$  represents a site of the periodic lattice, and therefore $\psi_n$ represents its \emph{unit cell}. The profile of the network is defined by the square matrices $H_0, H_1$. The transformation $\psi_n = x_n e^{-i E t}$ yields the eigenvalue problem associated to Eq.~\eqref{eq:fb_eq}, and then the  Bloch solution $x_n = e^{i {\mathbf k} n}y_k$ defined for the wave-vector ${\mathbf k}$ gives rise to the band structure $\{ E_j( {\mathbf k})\}_{j=1}^\nu$ of Eq.~\eqref{eq:fb_eq}. 

In this work we focus on ABF networks where all bands $E_j$ are independent on ${\mathbf k}$ -- hence all bands are flat. The collapse of the single-particle spectrum into several flatbands and the absence of dispersive states is coined caging. Any compact initial condition remains confined within a finite (compact) sub-volume of the network $\psi_n(t) \neq 0$ for $1\leq n\leq M$ and $\psi_n(t) = 0$ otherwise, for all $t\in\mathbb{R}$. 

In short-range flatband networks, the eigenstates associated to the flatband can always be recast as spatially compact.~\cite{read2017compactly} Using this as a starting point we prove the following:
\\
\\
{\it {\bf Theorem}: 
	Any one-dimensional $\nu\geq 2$ all bands flat network~\eqref{eq:fb_eq} with short-range hopping can be recast into a fully decoupled lattice 
	\begin{gather}
	    i \dot{\phi}_n =  H_R\phi_n \quad H_R = \text{diag}(E_1, E_2, \dots, E_\nu)
	    \label{eq:fb_eq_R}
	\end{gather}
	by applying a local detangling protocol $\psi_n \longmapsto \phi_n$ which nest a finite sequence of local unitary transformations each redefining the unit cell.
}
\\

The main idea behind this result is that in one dimension there always exist local unitary transformations each redefining the unit cells which recast all the flatband compact eigenstates within a single unit cell - revealing therefore their orthogonality. Therefore an equivalent statement of the theorem is that the $d=1$ ABF lattices always have orthonormal compact localized eigenstates. This statement is far from trivial: compact localised states of a flatband which coexists with other dispersive bands are not necessarily orthogonal -- \emph{e.g.} stub lattice~\cite{flach2014detangling,baboux2016bosonic,real2017flat} or the notable Lieb lattice~\cite{leykam2018artificial} provide counterexamples. Furthermore in the latter case compact states are not even complete (similarly to kagome and pyrochlore lattices~\cite{bergman2008band}). The detailed proof can be found in Appendix~\ref{app1a}. For the specific ABF case of nearest neigbour unit cell coupling in Eq.~\eqref{eq:fb_eq} it follows that compact localized states occupy two unit cells, and the detangling procedure involves one local unitary transformation which decouples the lattice into noninteracting $\nu$-mers. One subsequent unit cell redefinition and one more unitary transformation which diagonalizes the $\nu$-mer results in the detangled form~\eqref{eq:fb_eq_R}. Increasing the hopping range results in the corresponding increase of the number of nested unit cell redefinitions and local unitary transformations.

We conjecture that this result holds in higher dimensions, and any short-range ABF lattice in any dimensions is equivalent to decoupled sites up to a local unitary transformation. Note that the inverse is always true. Starting from a detangled set of sites, reversing the detangling procedure $\phi_n \longmapsto \psi_n$ yields a generator of any one dimensional ABF lattice, and of a plethora (if not any) of higher-dimensional ABF lattices, for any finite number of bands $\nu$. In the simplest case of nearest neighour unit cell coupling and a fixed unit cell redefinition one ABF manifold has dimension $2(\nu^2-1)$ since it is controlled by two unitary transformations (we removed the trivial global phases as irrelevant parameters). The manifold contains the detangled model. Since there are $2^{\nu}-2$ possible unit cell redefinitions, there are as many different ABF manifolds, all originating from one and the same detangled model. Extending to a nested sequence of $\mu$ unit cell redefinitions increases the hopping range to distance $\mu$ and dramatically increases the number of ABF manifolds as well as their dimensions. This scheme completes a list of several other generator schemes introduced in recent years~\cite{dias2015origami,morales2016simple,maimaiti2017compact,rontgen2018compact,toikka2018necessary,maimaiti2019universal} which instead were focused on flatband networks supporting both flat and dispersive bands.

The manifold of ABF lattices generated with this method include the known examples of Aharonov-Bohm caging, e.g. the Creutz ladder (as discussed below) and diamond chain in a fine-tuned magnetic  field.~\cite{vidal1998aharonov,doucot2002pairing,creutz1999end} On the other hand it is not at all obvious, and likely also not very relevant, whether any such generated ABF lattice is equivalent to some tight-binding lattice in a properly fine-tuned magnetic field, even if one allows for artificial fluxes in the spirit of the Haldane model.~\cite{haldane1988model}

\begin{figure*}
    \centering
    \includegraphics[width=2\columnwidth]{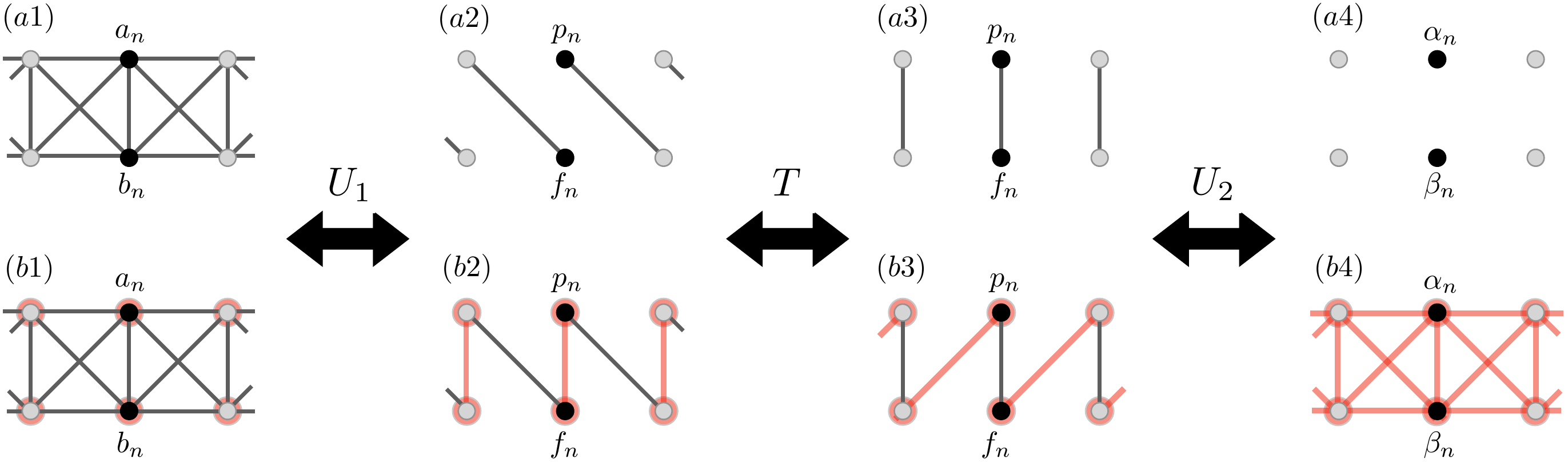}
    \caption{Schematic representation of the unit cell redefinition for a $\nu=2$ ABF lattice. In each panel, the black dots label the chosen unit cell. The solid gray lines represent the linear hopping terms; the red shaded lines represent the interaction terms. (a1)-(a4) Noninteracting regime. (b1)-(b4) Interacting regime.}
    \label{fig:abf_rot}
\end{figure*}

Let us visualize this procedure for the simplest case of $\nu=2$ networks in Fig.~\ref{fig:abf_rot}(a1)-(a4), with the canonical coordinates $\psi_n = (a_n, b_n)$ and the detangled coordinates $\phi_n=(\alpha_n, \beta_n)$. The detangling procedure $\psi_n \longmapsto \phi_n$ unfolds in three steps
\begin{gather}
    \begin{pmatrix}
        a_n  \\[0.3em]
        b_n 
    \end{pmatrix}
    \xmapsto{\ \ U_1\ \ }
    \begin{pmatrix}
        p_n  \\[0.3em]
        f_n 
    \end{pmatrix}
    \xmapsto{\ \ T\ \ }
    \begin{pmatrix}
        p_n  \\[0.3em]
        f_n 
    \end{pmatrix}
    \xmapsto{\ \  U_2\ \  }
    \begin{pmatrix}
        \alpha_n  \\[0.3em]
        \beta_n 
    \end{pmatrix}
    \label{eq:par_nu2}
\end{gather}
with the alternation of two unitary transformations $U_1$ and $U_2$  and one relabeling of the lattice sites $T$:
\begin{gather}
    U_i 
     = e^{i\theta_i}
    \begin{pmatrix}
        z_i &  w_i   \\[0.3em]
        -  w^*_i & z^*_i
    \end{pmatrix}
    \quad \quad 
    T:\  \begin{cases}
        p_n \longmapsto p_n\\
        f_n \longmapsto f_{n-1}
    \end{cases}
    \label{eq:U12}
\end{gather}
The complex numbers $z_i,w_i$ are constrained with $|z_i|^2 + |w_i|^2 = 1$. Without loss of generality the two flatband energies can be locked at $E=\pm1$. We then parametrize the matrices $H_0,H_1$ for $\nu=2$ ABF networks in Eq.~\eqref{eq:fb_eq} as
\begin{align}
    \label{eq:H0H1_rot0}
    H_0 &= \Gamma_0
    \begin{pmatrix}
        |z_1|^2 - |w_1|^2 & -2 z_1 w_1 \\[0.3em]
        -2 z_1^* w_1^* & |w_1|^2 - |z_1|^2
    \end{pmatrix}\\ 
    &H_1 = \Gamma_1
    \begin{pmatrix}
        z_1 w_1^* & z_1^2 \\[0.3em]
        - (w_1^*)^2 & -z_1 w_1^*
    \end{pmatrix}
    \label{eq:H0H1_rot1}
\end{align}
with $\Gamma_0 = |w_2|^2 - |z_2|^2$ and $\Gamma_1 = 2 z_2 w_2$ (see Appendix~\ref{app1b} for details). The resulting manifold of $\nu=2$ ABF lattices in Eqs.~(\ref{eq:H0H1_rot0}-\ref{eq:H0H1_rot1}) includes the notable Creutz lattice,~\cite{creutz1999end}, which is obtained for $z_1 = z_2 = w_2 = 1/\sqrt{2}$ and $w_1 = i/\sqrt{2}$. Our manifold also includes a lower-dimensional submanifold of lattices related to the Creutz lattice via a gauge transformation.~\cite{tovmasyan2018preformed}~\footnote{Consider a gauge transformation $\mathcal{G}: \{a_n,b_n\} \longmapsto \{e^{i\xi_n} a_n, e^{i\eta_n} b_n\}$ defined for arbitrary phase functions $\xi_n,\eta_n$. The submanifold of lattices related to the Creutz ladder via $\mathcal{G}$ is given by the family of parameters
$z_1 = e^{i\frac{\xi_n-\eta_n}{2}}/\sqrt{2}, \   
w_1 =  i e^{i\frac{\xi_n-\eta_n}{2}} /\sqrt{2}, \  
z_2 =  e^{i g_n(\xi_n , \eta_n)}/\sqrt{2},\  
w_2 =  e^{-i g_n(\xi_n, \eta_n)}/\sqrt{2}$ 
in Eqs.~(\ref{eq:H0H1_rot0}-\ref{eq:H0H1_rot1}) for arbitrary functions $g_n$.}

The result that any 1D ABF lattice is unitarily equivalent to a set of decoupled sites provides a powerful framework for analysis of ABF networks, for example their transport properties, in presence of various perturbations, in particular interactions/nonlinearities.

\section{Nonlinear interactions: sub-diffusion and fine-tuned caging}
\label{sec:nl-int}

Any linear ABF network cages any localized initial excitation. An important question is the fate of this caging behaviour in  presence of interactions. We will address this question in the following way: A given linear ABF Hamiltonian is a member of a manifold of ABF Hamiltonians linked together by (local) unitary transformations. We pick one of the members of that manifold, add local nonlinear interactions, and need to figure whether caging is destroyed or not. For that we transform the chosen member into the detangled basis, inspect the transformed nonlinear interactions and arrive at necessary and sufficient conditions for nonlinear caging.

Let us add nonlinear terms to the Schr\"odinger equation~\eqref{eq:fb_eq} which result e.g. from a mean-field approximation to a bosonic many-body interacting system. For convenience we choose the local Kerr-like nonlinearity. This choice is not essential for the following arguments, and is made for convenience only. Equation~\eqref{eq:fb_eq} turns into 
\begin{gather}
    \label{eq:fb_NL}
    i \dot{\psi}_n = - H_0\psi_n -  H_1\psi_{n+1} - H_1^\dagger\psi_{n-1} + U F(\vert\psi_n\vert^2) \psi_n.
\end{gather}
Here $F(\vert\psi_n\vert^2)$ is a diagonal matrix with nonzero elements $F_{\mu,\mu} \equiv \vert\psi_{n,\mu} \vert^2$. The above Gross-Pitaevski-type lattice equations are generated by the Hamiltonian:
\begin{align}
    i\dot{\psi}_n &= \nabla_{\psi_n^{\ast}}\mh_{G},\qquad \hmh_{G} = \hmh_0^{G} + \hmh_1^{G}\\
    \label{eq:dnls_ham1}
    \mh_0^{G} &= - \sum_{n\in\mathbb{Z}} \left[ \frac{1}{2}\left(\psi_{n}^{* T} H_0\psi_n \right) + \left(\psi_{n}^{* T} H_1\psi_{n+1} \right) + \text{h.c.}\right],\\
    \label{eq:dnls_ham_NL}
    \mh_1^G &= \frac{U}{2}\sum_{n\in\mathbb{Z}} F^2(\vert\psi_n\vert^2).
\end{align}
The coordinate redefinition $\psi_n \longmapsto \phi_n$ to Eq.~\eqref{eq:fb_NL} decouples the quadratic part. The local nonlinear terms in the original representation $\psi_n$ turn nonlocal in the new coordinates $\phi_n$ of the detangled basis. The evolution equation~\eqref{eq:fb_NL} in the new representation reads 
\begin{gather}
    i \dot{\phi}_n =  H_R\phi_n + g\mathcal{P}(\{\phi_n\}).
    \label{eq:fb_eq_R_NL}
\end{gather}
$H_R$ is the diagonal matrix in Eq.~\eqref{eq:fb_eq_R} and $\mathcal{P}$ a homogeneous polynomial of degree three in $\{\phi_n\}$:
\begin{gather}
    \mathcal{P}(\{\phi_n\}) = \sum_{mb;kc,ld} V_{na,mb;kc,ld}\phi_{m,b}^*\phi_{k,c}\phi_{l,d}.
\end{gather}
Here $n$ labels unit cells and $a$ labels sites in the unit cell $n$. The matrix elements $V_{na,mb;kc,ld}$ define a nonlinear interaction network in the detangled basis with the corresponding Hamiltonian
\begin{gather}
    \label{eq:gen-h1}
    \mh_1^G = \frac{U}{2} \sum_{na,mb;kc,ld} V_{na,mb;kc,ld}\phi_{n,a}^*\phi_{m,b}^*\phi_{k,c}\phi_{l,d}.
\end{gather}
Note that it is straightforward to consider e.g. a two-dimensional square lattice or three-dimensional cubic lattice, which  leaves the expressions~\eqref{eq:gen-h1} invariant while turning the unit cell indices $n,m$ into two-component or three component vectors with integer components, respectively.

\subsection{Necessary and sufficient condition\\ for nonlinear caging}
\label{sec:iff-caging}

Nonlinear caging is defined similar to the linear case as confinement of initial excitation
\begin{gather}
    \begin{cases}
        \psi_n(t) \neq 0 \quad |n|\leq M\\
        \psi_n(t) = 0 \quad |n| > M
    \end{cases}
    \qquad\forall\,t > 0
    \label{eq:caging}
\end{gather}
For convenience we refer to the set of nonzero amplitudes as an \emph{excitation}. To study the spreading of this initial condition it is enough to consider sites $n,a$ in the unit cell just outside the excitation: if the amplitude on those sites becomes nonzero, by induction it follows that the spreading will continue to other unit cells outside the initial excitation.

The time evolution of the amplitude on a site $n,a$ outside the initial excitation is governed by:
\begin{gather}
    \label{eq:gen-ue}
    i\dot{\phi}_{n,a} = E_a\phi_{n,a} + \sum_{mb;kc,ld} V_{na,mb;kc,ld} \phi_{m,b}^*\phi_{k,c}\phi_{l,d},\\
    \phi_{n,a}(t=0) = 0. \notag
\end{gather}
At time $t=0$ it follows by assumption that $\phi_{n,a}=0$. Nonlinear caging implies that $\phi_{n,a}=0$ for all times. This can happen if and only if the total nonlinear contribution on the rhs of (\ref{eq:gen-ue}) vanishes at all times. 
Excluding accidental cancellations of the time-dependent rhs terms, the caging requirement is equivalent to enforcing the vanishing of each individual term in the above sum.~\footnote{One could potentially imagine further fine-tuning of the interaction that might cancel the total sum but not the individual terms: this is however only possible for a specific combination of the amplitudes $\phi_{n,a}$, implying that at best only specific excitations can be caged. We are interested in caging of any initial excitation and therefore do not investigate further this possibility.} Therefore at least one amplitude in all the terms $\phi_{m,b}^*\phi_{k,c}\phi_{l,d}$ must be located outside the original excitation (and be therefore zero as well). Furthermore if that one amplitude is located in a unit cell different from $n$ in at least one interaction term, we can use translation invariance and shift the interaction, so that all the amplitudes in that interaction term are located inside the excitation.~\footnote{We are relying here on the requirement that any initial excitations are caged, so that we do not need to worry about some particularly small excitations, where this argument might fail.} Then the interaction term becomes nonzero and will induce spreading of the initial excitation. 
Therefore at least $m=n$, or $k=n$, or $l=n$. However because of the Hermiticity and translation invariance of the problem the remaining two amplitudes have also to be equal, otherwise we can find a unit cell outside the excitation, with a nonzero nonlinear term in the evolution equation. Therefore all the unit cells in the amplitudes $\phi$ of the interaction terms in Eq.~\eqref{eq:gen-h1} have to appear in pairs, so that Eq.~\eqref{eq:gen-ue} becomes:
\begin{align}
    \label{eq:gen-ue-caged}
    i\dot{\phi}_{n,a} = E_a\phi_{n,a} + & \sum_{b;mc,d} V_{na,nb;mc,md} \phi_{n,b}^*\phi_{m,c}\phi_{m,d}\\
    + & \sum_{b;mc,d} V_{na,mb;nc,md} \phi_{m,b}^*\phi_{n,c}\phi_{m,d}, \notag\\
    \phi_{n,a}(t=0) = & 0. \notag
\end{align}
The corresponding Hamiltonian becomes
\begin{align}
    \label{eq:gen-ham-caged}
    \mh_1^G =  & \sum_{na,b;mc,d} V_{na,nb;mc,md} \phi_{n,a}^*\phi_{n,b}^*\phi_{m,c}\phi_{m,d}\\
    + & \sum_{na,b;mc,d} V_{na,mb;nc,md} \phi_{n,a}^*\phi_{m,b}^*\phi_{n,c}\phi_{m,d}. \notag
\end{align}
To achieve caging, it is necesessary and sufficient if a given Hamiltonian with a linear ABF part can be transformed into Eq.~\eqref{eq:gen-ham-caged} in the detangled basis. 

Our approach can be readily extended to more complicated interactions, e.g. involving higher powers of densities or nonlocal interactions (as has been pointed out already in Ref.~\onlinecite{diliberto2019nonlinear}), as well as to higher lattice dimensions, whenever single-particle Hamiltonian admits detangling,~\cite{flach2014detangling} which we conjecture to be true for any ABF Hamiltonian. We arrived at a direct way to test for caging in a given nonlinear AFB network with local nonlinearity: transform into the detangled basis,  obtain the transformed interaction Hamiltonian, and check that all the nonlinear terms are included in Eq.~\eqref{eq:gen-ham-caged}. If they are, the nonlinear network exhibits caging, otherwise it does not.

Now we can address the question whether a given linear ABF manifold contains a submanifold which supports nonlinear caging when adding local Kerr-like nonlinearities. We remind that the ABF manifold is supposed to contain the detangled ABF Hamiltonian member. The manifold members are connected by the action of a pair of unitary transformations. Each unitary transformation is controlled by $\nu^2$ parameters. The nonlinear caging reduces to zeroing a number of \emph{dangerous} terms in the transformed nonlinearity, which amounts to the same number of equations for the unitary transformation parameters. Let us zero one element in the unitary transformation $U_1$. That leads to a zeroing of $\nu^3$ nonlinear coefficients $V$ in Eq.~\eqref{eq:gen-h1}. It therefore appears that we can always remove all non-caging terms and remain with a non-empty submanifold of nonlinear caging Hamiltonians.

\subsection{Two band networks}
\label{sec:2bands}

We now illustrate the above generic result for nonlinear caging with examples drawn from the fully parametrized class of $\nu=2$ ABF networks~(\ref{eq:H0H1_rot0}-\ref{eq:H0H1_rot1}). 
Equation~\eqref{eq:fb_NL} with $\psi_n = (a_n,b_n)$ results in
\begin{gather}
    \label{eq:fb_NL_nu2}
    i \dot{\psi}_n = - H_0\psi_n -  H_1\psi_{n+1} - H_1^\dagger\psi_{n-1} + U F(\vert\psi_n\vert^2 ) \psi_n,\\
    \text{with} \;\; F(\vert\psi_n\vert^2) = 
    \begin{pmatrix}
        |a_n|^2 & 0 \\[0.3em]
        0 & |b_n|^2
    \end{pmatrix},\notag
\end{gather}
while the nonlinear Hamiltonian $\mh_1^G$ reads
\begin{align}
    \mh_1^G &= \frac{U}{2}\sum_{n\in\mathbb{Z}}\left[ |a_{n}|^4  + |b_{n}|^{4} \right].
    \label{eq:dnls_ham_NL_nu2}
\end{align}
The caging condition can be established already in the $p_n, f_n$ representation (see Fig.~\ref{fig:abf_rot}). Indeed the transformation $U_2$ (see Fig.~\ref{fig:abf_examples_tevo}) only affects the couplings inside one unit cell and therefore cannot introduce terms that violate the nonlinear caging criterion. This results in the following necessary and sufficient condition (see Appendix~\ref{app2c} for details):
\begin{gather}
    \label{eq:fine_tuning}
    |w_1|^2 = |z_1|^2.
\end{gather}   
Then the Hamiltonian $\mh_1^G$ in Eq.~\eqref{eq:dnls_ham_NL} recast via the transformations $U_1$ and $T$ in Eqs.~(\ref{eq:par_nu2}-\ref{eq:U12}) is represented by the amplitudes $(p_{n}, f_{n})$ (see Appendix~\ref{app3a})
\begin{align}
    \mh_1^G & = U\sum_n \left\{ |z_1|^4 \left[ |p_n|^4 + |f_n|^4 + 4 |p_{n}|^2 |f_{n+1}|^2 \right] \right. \notag \\
    \label{eq:Ham1_GP_R}
    & \left. \qquad\quad + z_1^{*2} w_1^2  p_n^{*2} f_{n+1}^2 + z_1 w_1^{*2} p_n^{2} f_{n+1}^{*2} \right\}.
\end{align}
The full Hamiltonian in the detangled representation is obtained in Appendix~\ref{app:fd-ab} for two cases - one which satisfies caging, and another which does not. The condition~\eqref{eq:fine_tuning} leads to a fine-tuned subclass of nonlinear lattices which support nonlinear caging.

\subsubsection{Two examples}
\label{sec:NL_ex}

We test two example networks parametrized via Eqs.~(\ref{eq:H0H1_rot0},\ref{eq:H0H1_rot1}) -- one which satisfies Eq.~\eqref{eq:fine_tuning} and one which does not -- generated by setting  $z_i = \cos \varphi_i$, $w_i = \sin \varphi_i$ in Eqs.~(\ref{eq:H0H1_rot0},\ref{eq:H0H1_rot1}). 

\begin{figure}
    \centering
    \includegraphics[width=0.9\columnwidth]{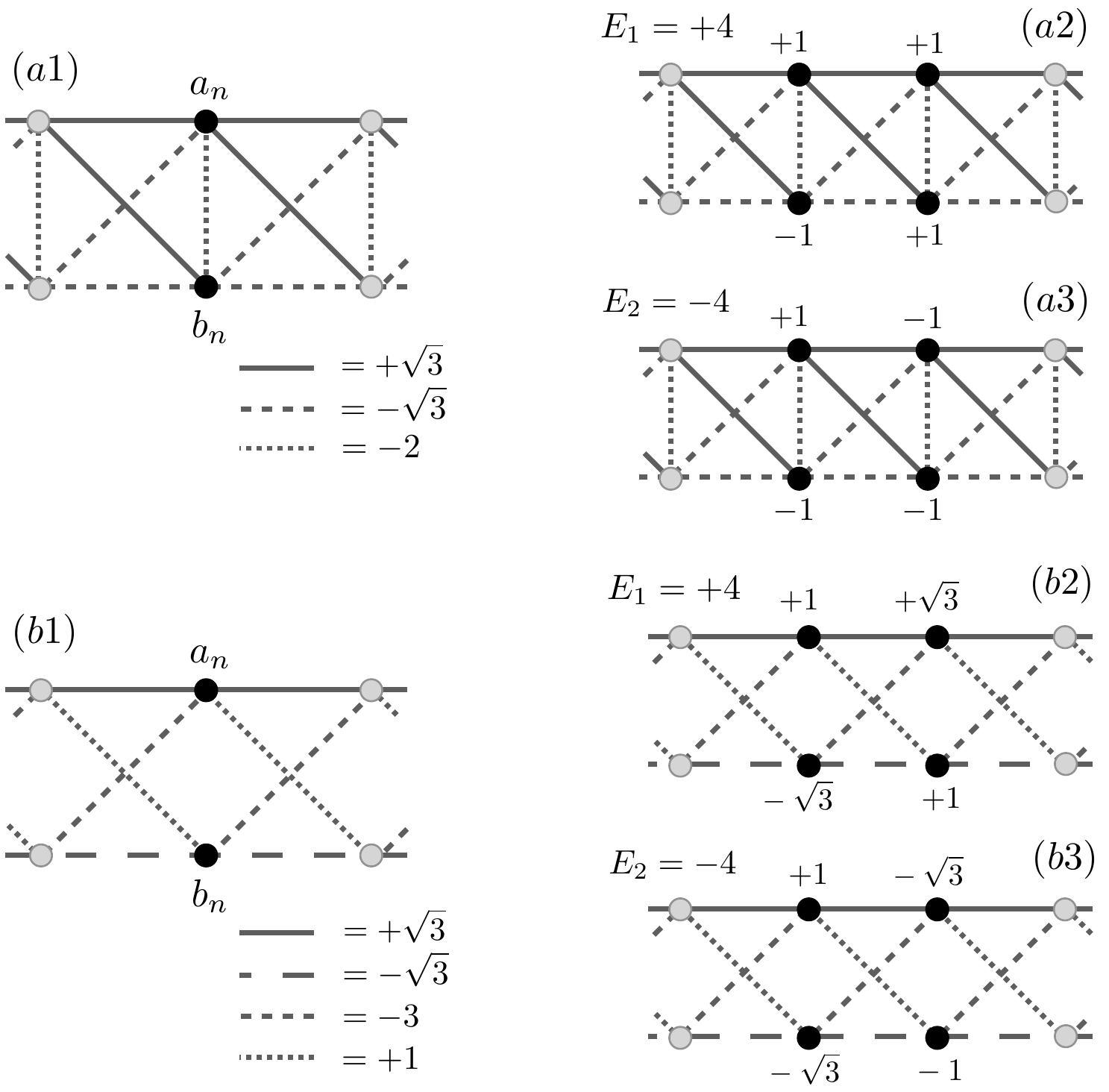}
    \caption{(a1) Schematic representation of model A, Eq.~\eqref{eq:ex1}. (a2)-(a3) CLSs of the two flatbands at $E_{1,2} = \pm 4$. (b1) Schematic representation of model B, Eq.~\eqref{eq:ex2}. (b2)-(b3) CLSs of the two flatbands at $E_{1,2} = \pm 4$.}
    \label{fig:abf_examples}
\end{figure}

\emph{Model A} is obtained by setting $\varphi_1 = \pi/4$ and $\varphi_2 = \pi/6$ and satisfies nonlinear caging Eq.~\eqref{eq:fine_tuning}: 
\begin{align}
    i\dot{a}_n &=  -2 b_n + \sqrt{3}\left(a_{n+1} + a_{n-1} + b_{n+1} - b_{n-1}\right) + U a_n|a_n|^2, \notag \\
    i\dot{b}_n &=  -2 a_n - \sqrt{3}\left( b_{n+1} + b_{n-1} + a_{n+1} - a_{n-1}\right)  + U b_n|b_n|^2.
    \label{eq:ex1}
\end{align}
The network schematics is shown in Fig.~\ref{fig:abf_examples}(a1). For $U=0$, there are two flatbands at $E_{1,2} = \pm 4$ with the respective CLSs shown in Fig.~\ref{fig:abf_examples}(a2)-(a3). 
Notably model A cannot be obtained from the previously studied Creutz ladder nor from its gauge transformation related partners since $\varphi_1 \neq \varphi_2$.

\emph{Model B} is obtained by setting $\varphi_1 = \pi/6$ and $\varphi_2 = \pi/4$. This model does not provide nonlinear caging since the fine-tuning condition~\eqref{eq:fine_tuning} is violated. The equations read 
\begin{align}
    i\dot{a}_n &= \sqrt{3} a_{n+1} + \sqrt{3} a_{n-1} + b_{n+1} - 3 b_{n-1} + U a_n|a_n|^2, \notag \\
    i\dot{b}_n &= -\sqrt{3} b_{n+1} - \sqrt{3} b_{n-1} - 3 a_{n+1} + a_{n-1} + U b_n|b_n|^2.
    \label{eq:ex2}
\end{align}
The network schematics is shown in Fig.~\ref{fig:abf_examples}(b1). For $U=0$, there are two flatbands at $E_{1,2} = \pm 4$ with the respective  CLSs shown in Fig.~\ref{fig:abf_examples}(b2)-(b3). 

\begin{figure}
    \centering
    \includegraphics[width=0.925\columnwidth]{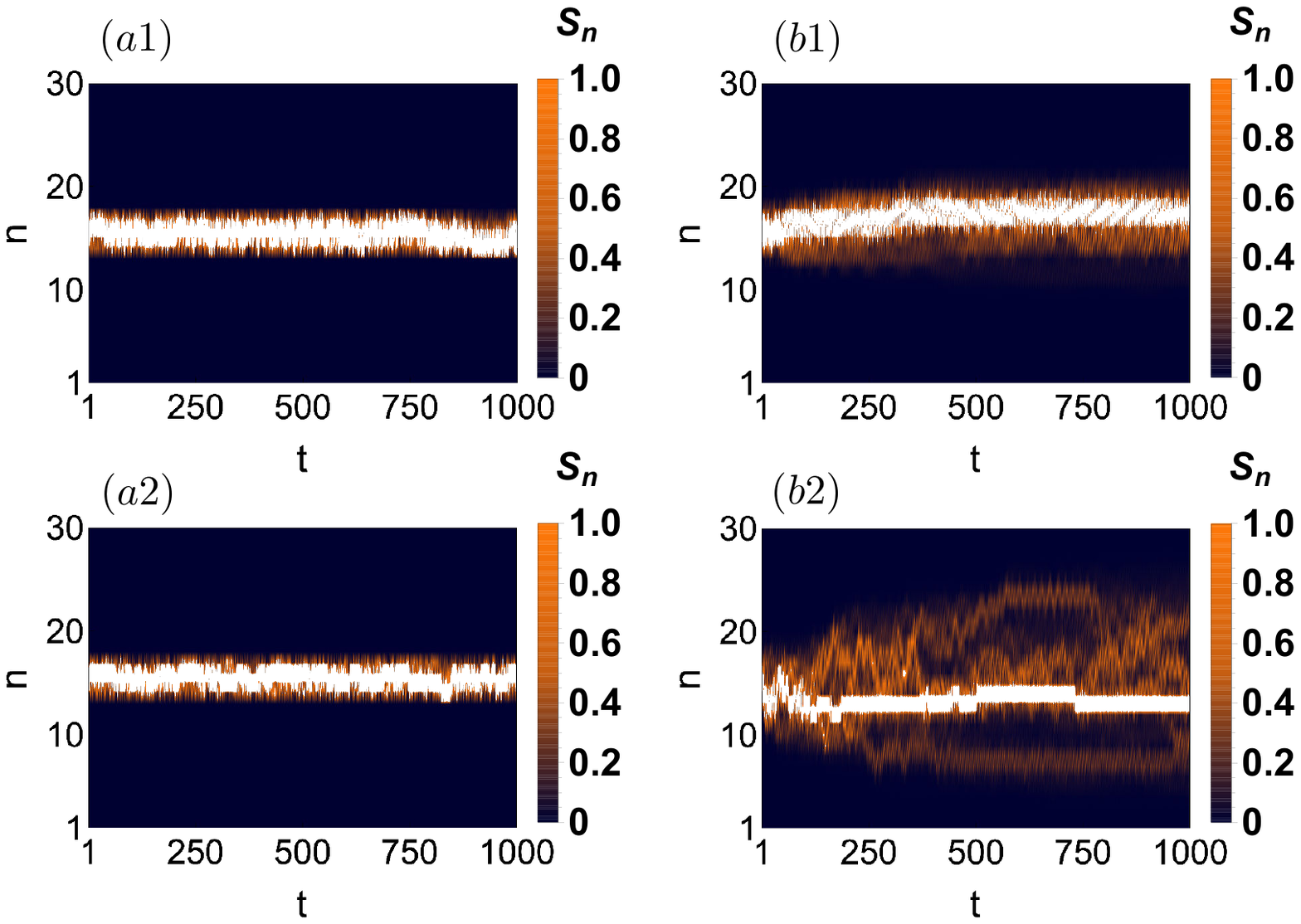}
    \caption{(a1)-(a2) time-evolution of $\text{IC}_1$ according to model A, Eq.~\eqref{eq:ex1} for $U=1$ and $U=5$ respectively. (b1)-(b2)  same as (a1)-(a2) for model B, Eq.~\eqref{eq:ex2}.}
    \label{fig:abf_examples_tevo}
\end{figure}

We visualize the presence (respectively absence) of nonlinear caging in these models by numerically computing the time evolution of an initially compact localized excitation.
We use second order splitting ABC schemes for the numerical integration.~\cite{danieli2019computational} We consider a sample compact excitation $\text{IC}_1$ spanning over two unit cells, and we evolve the local density  $S_n = |a_n|^2 + |b_n|^2$. The results are shown in Fig.~\ref{fig:abf_examples_tevo} for both models A~\eqref{eq:ex1} and B~\eqref{eq:ex2} and for two interaction strengths $U=1$ and $U=5$. For model A, Eq.~\eqref{eq:ex1} -- panels (a1)-(a2) -- the initial compact excitation  $\text{IC}_1$ remains confined within four unit cells, confirming the expected nonlinear caging. For model B, Eq.~\eqref{eq:ex2} -- panels (b1)-(b2) -- the initial excitation is propagating into the chain, confirming that caging is lost.

\begin{figure}
    \centering
    \includegraphics[width=0.85\columnwidth]{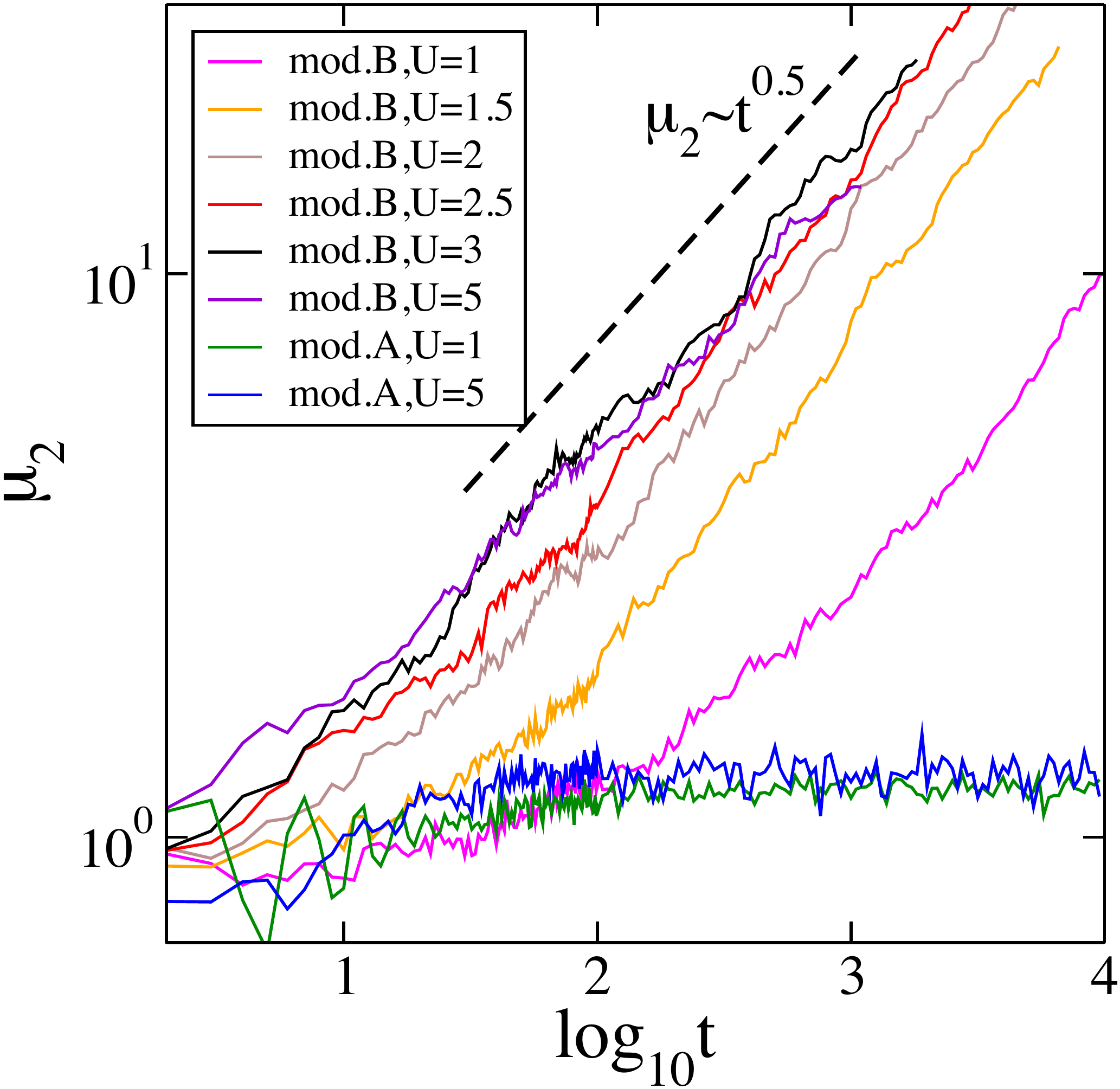}
    \caption{Time-evolution of the second moment $\mu_2$ over an ensemble of $48$ initial conditions according to model B, Eq.~\eqref{eq:ex2} (upper five curves) and model A, Eq.~\eqref{eq:ex1} (bottom two curves) with $N=40$ for different $U$.}
    \label{fig:ex2_tevo_m2}
\end{figure}

In Fig.~\ref{fig:ex2_tevo_m2} we show the time evolution of the second moment $\mu_2$ defined as
\begin{align}
    \label{eq:obs_m2}
    \mu_2 &= \sum_{n=1}^{N} [(X - n)^2 (|a_n|^2 + |b_n|^2 )]
\end{align}
with $X=\sum_{n=1}^{N} [ n (|a_n|^2 + |b_n|^2 )]$ for both models. The curves have been averaged over an ensemble of $48$ compact initial conditions spanning over two unit cells all chosen with the same total norm $S = \sum_n S_n=7$. In the case of the cage-preserving model A, Eq.~\eqref{eq:ex1} we observe no signature of spreading as the second moment remains $\mu_2\sim 1$ over time. In the case of the non-caging preserving model B, Eq.~\eqref{eq:ex2} we observe a subdiffusive spreading regime: within the studied time-window our data agree semi-quantitatively with $\mu_2\sim t^{0.5}$ for various values of the interaction strength $U$. The details of this process and its relation to previous studies of nonlinear destruction of Anderson localization~\cite{laptyeva2014nonlinear} is certainly an interesting future project. We conjecture here that subdiffusion results from weak interactions renormalizing the compact localized states and inducing nonlinear interactions between them. Both effects are proportional to the local norm density which decreases with further spreading of the wave packet. 

\subsection{Generalizations to more bands and higher lattice dimensions}
\label{seec:nu34_D2}

We now make use of the approach introduced in Sec.~\ref{sec:iff-caging} and employed in Sec.~\ref{sec:2bands} for one-dimensional $\nu=2$ lattices to present cage preserving nonlinear lattices with $\nu\geq 3$ and higher lattice dimensions.

\begin{figure}
    \centering
    \includegraphics[width=0.85\columnwidth]{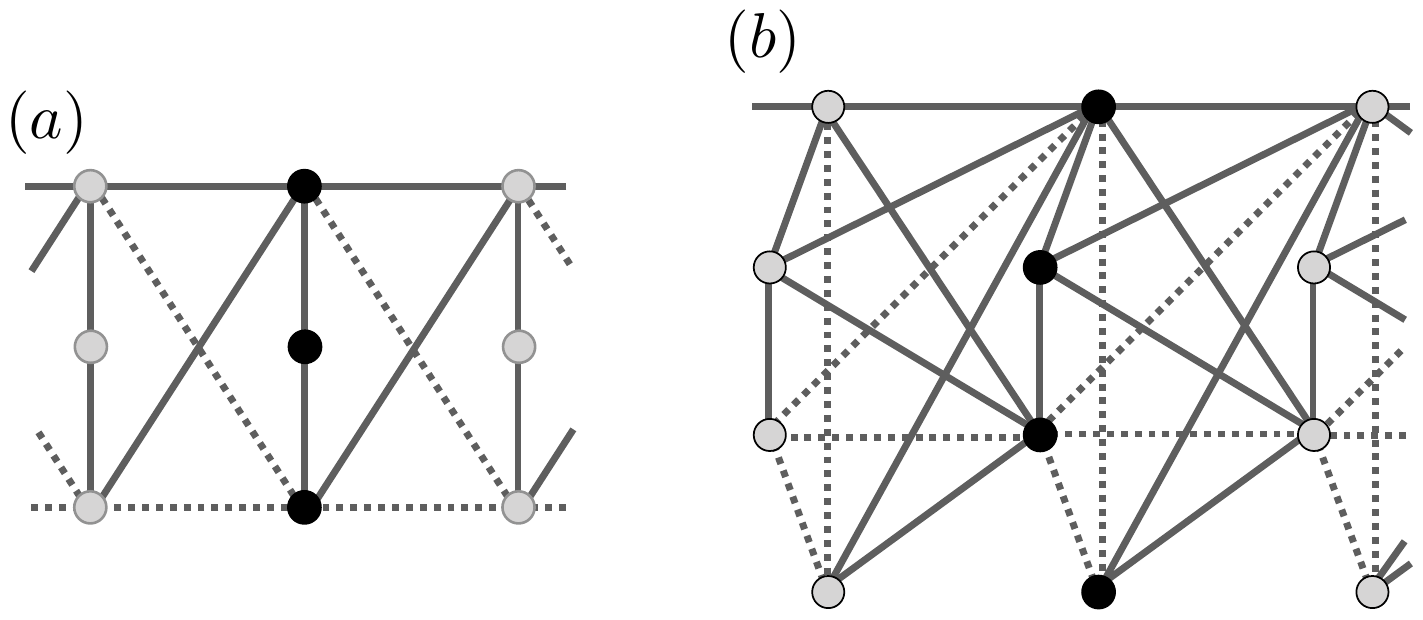}
    \caption{Sample $\nu=3$ (a) and $\nu=4$ (b) ABF lattices defined in Eq.~\eqref{eq:H0H1_1D_nu3} and in Eq.~\eqref{eq:H0H1_1D_nu4} respectively. For sake of simplicity, in these drawings all parameters $\mu,\chi$ have been set to 0.  Solid lines: hopping amplitude $t=+1$. Dashed lines; hopping amplitude $t=-1$. These lattices preserve nonlinear caging.}
    \label{fig:caging_nu3_4}
\end{figure}

In one-dimension, we consider the family of $\nu=3$ ABF lattices Eq.~\eqref{eq:fb_NL} defined by 
\begin{gather}
    H_0 = \begin{pmatrix}
        \mu & 1 & -\mu \\[0.3em]
        1 & 0 & 1 \\[0.3em]
        -\mu & 1 & \mu 
    \end{pmatrix},\quad 
    H_1 = \begin{pmatrix}
        1 & \chi & 1 \\[0.3em]
        0 & 0 & 0 \\[0.3em]
        -1 & -\chi & -1 
    \end{pmatrix}
    \label{eq:H0H1_1D_nu3}
\end{gather}
with free parameters $\mu,\chi$. The family -- shown in Fig.~\ref{fig:caging_nu3_4}(a) for $\mu,\chi=0$ for clarity -- preserves nonlinear caging since the unitary transformation that detangles the linear part 
\begin{gather}
    U_{\nu=3} = \frac{1}{\sqrt{2}}
    \begin{pmatrix}
        1  &  0 &  1  \\[0.3em]
        0 &  \sqrt{2} & 0  \\[0.3em]
        -1  & 0 & 1 
    \end{pmatrix},
    \label{eq:tr_1D_nu3}
\end{gather}
does not generate any forbidden transporting nonlinear terms in the detangled representation. The $\nu=3$ ABF family in Eq.~\eqref{eq:H0H1_1D_nu3} is a submanifold of the full $\nu=3$ ABF manifold. The full manifold also contains the ABF diamond chain which preserves nonlinear caging as well as studied in Refs.~\onlinecite{gligoric2018nonlinear,diliberto2019nonlinear}, but is not part of the above example family. This demonstrates that previously observed ABF networks which satisfy nonlinear caging are single members of entire families of multi-parameter ABF submanifolds which preserve nonlinear caging.

The very same reasoning applies to larger number of bands, \emph{e.g.} for $\nu=4$. The family of nonlinear lattices Eq.~\eqref{eq:fb_NL} defined by
\begin{equation}
    \small 
    H_0 = \begin{pmatrix}
        \mu_1  &  1 & -\mu_1 & -1  \\[0.3em]
        1 &  \mu_2 & 1 & \mu_2  \\[0.3em]
        -\mu_1  &  1 &  \mu_1 & -1  \\[0.3em]
        -1 &  \mu_2 & -1 & \mu_2
    \end{pmatrix},\quad 
    H_1 = \begin{pmatrix}
        1  &  \chi_1 &  1 & -\chi_1  \\[0.3em]
        1  &  \chi_2 &  1 & -\chi_2 \\[0.3em]
        -1  &  -\chi_1 & - 1 & \chi_1  \\[0.3em]
        1  &  \chi_2 &  1 & -\chi_2 
    \end{pmatrix}
    \label{eq:H0H1_1D_nu4}
\end{equation}
with free parameters $\mu_1,\mu_2,\chi_1,\chi_2$ -- shown in Fig.~\ref{fig:caging_nu3_4}(b) with $\mu_1,\mu_2,\chi_1,\chi_2=0$ for clarity -- is cage preserving. Indeed the unitary transformation that detangles the linear part 
\begin{equation}
    U_{\nu=4} = \frac{1}{\sqrt{2}}
    \begin{pmatrix}
        1  &  0 &  1 &0 \\[0.3em]
        0 &  1 & 0 & 1 \\[0.3em]
        -1  & 0 & 1  & 0 \\[0.3em]
        0 &  -1 & 0 & 1 
    \end{pmatrix}
    \label{eq:tr_1D_nu4}
\end{equation}
does not introduce any \emph{dangerous} transporting nonlinear terms in the detangled representation.

\begin{figure}
    \centering
    \includegraphics[width=0.95\columnwidth]{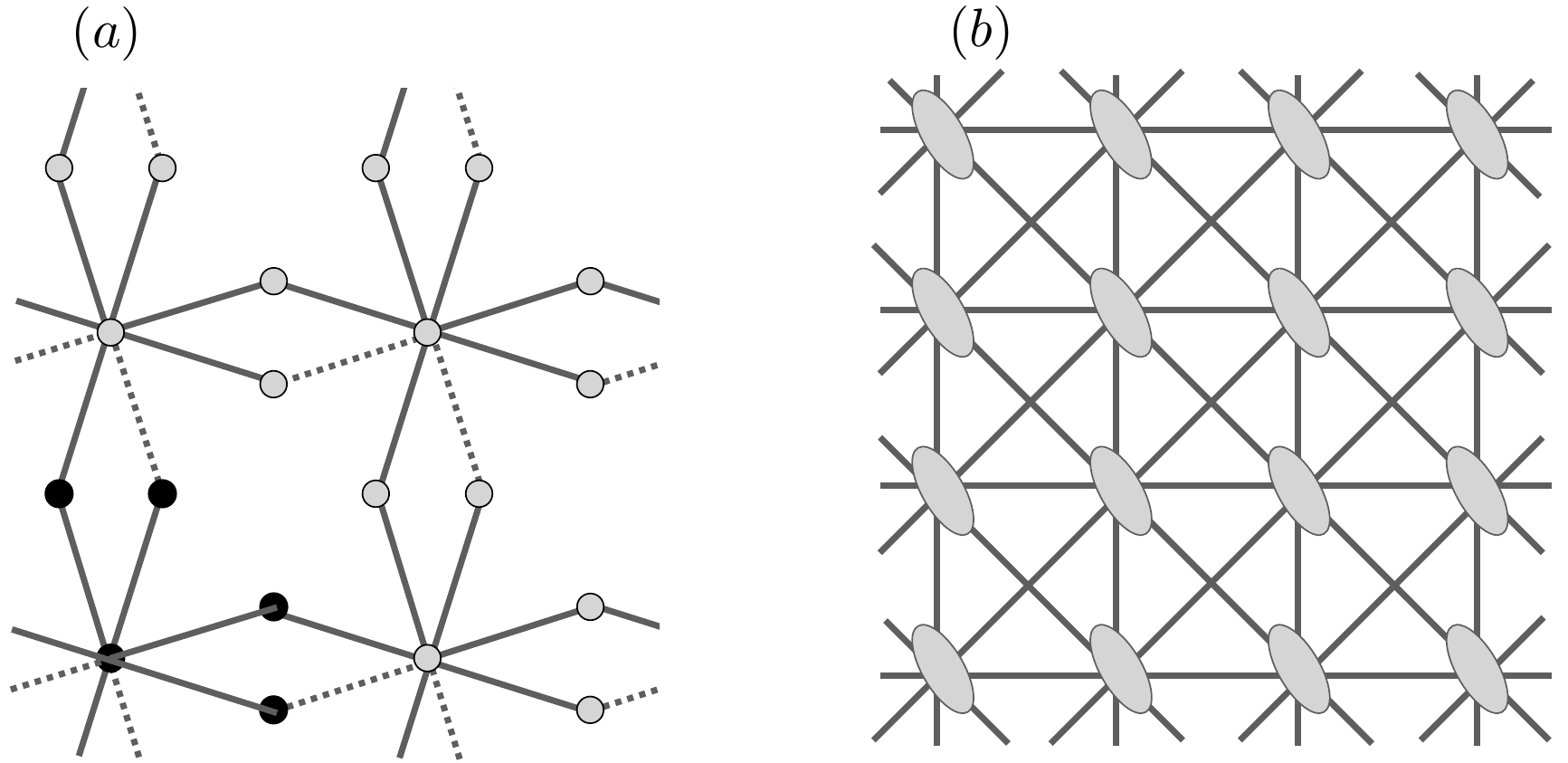}
    \caption{Nonlinear caging in two dimensions. (a) $\nu=5$ cage preserving two-dimensional (decorated Lieb) lattice. Solid lines indicate hopping $+1$, dashed lines indicate hopping $-1$. (b) $\nu=2$ cage preserving two-dimensional lattice with second nearest neighbor  hopping defined in Eqs.~(\ref{eq:fb_2D_nu2},\ref{eq:H0H1_2D_nu2}). For sake of clarity, the tilted ellipses represent the unit cells, while the solid lines represent the hopping matrix elements Eq.~\eqref{eq:H0H1_2D_nu2}.}
    \label{fig:caging_2D}
\end{figure}

In two dimensions, we use the $\nu=5$ two dimensional lattice called \emph{decorated Lieb lattice}~\cite{rontgen2019quantum} as one example which satisfies nonlinear caging (see Fig.~\ref{fig:caging_2D}(a)).

Another example is a novel two-dimensional $\nu=2$ ABF lattice with additional diagonal hopping connectivities (see Fig.~\ref{fig:caging_2D}(b)):
\begin{equation}
    \begin{split}
        i \dot{\psi}_{n,m} = & -H_{1,n}\psi_{n+1,m} - H_{1,n}^\dagger\psi_{n-1,m} \\
        & -H_{1,m}\psi_{n,m+1} - H_{1,m}^\dagger\psi_{n,m-1} \\
        & -H_{2,D}\psi_{n+1,m+1} - H_{2,D}^\dagger\psi_{n-1,m-1} \\
        & -H_{2,A}\psi_{n+1,m-1} - H_{2,A}^\dagger\psi_{n-1,m+1} \\
        & +U F(\vert\psi_{n,m}\vert^2) \psi_{n,m}
    \end{split}
    \label{eq:fb_2D_nu2}
\end{equation}
with
\begin{equation}
    \begin{split}
        H_{1,n} &= - \frac{\sqrt{3}}{4}
        \begin{pmatrix}
            0 & 1 \\[0.3em]
            1 & 0 
        \end{pmatrix},\ \qquad\ 
        H_{1,m} = \frac{1}{4}
        \begin{pmatrix}
            -1 & -1 \\[0.3em]
            1 & 1 
        \end{pmatrix} \\
        H_{2,D} &= \frac{\sqrt{3}}{8}
        \begin{pmatrix}
            1 & 1 \\[0.3em]
            -1 & -1 
        \end{pmatrix},\quad
        H_{2,A} =  \frac{\sqrt{3}}{8}
        \begin{pmatrix}
            -1 & 1 \\[0.3em]
            -1 & 1 
        \end{pmatrix}  
    \end{split}   
    \label{eq:H0H1_2D_nu2}
\end{equation}
This model is cage preserving since the detangling process obtained by the unitary transformation
\begin{gather*}
    U_{\nu=2} = \frac{1}{\sqrt{2}}
    \begin{pmatrix}
        1 & 1 \\
        -1 & 1  
    \end{pmatrix}
\end{gather*}
twice alternated by the redefinition $\psi_{n,m} \longmapsto \psi_{n,m-1}$ reduces the Kerr nonlinearity to the form given by Eq.~\eqref{eq:gen-ham-caged}.

\section{Discussions and perspectives}

In this work we showed that in one dimension dispersionless networks can be completely detangled via local unitary transformations. The inversion of that procedure yields a systematic generator for all band flat networks with finite-range hopping terms in any lattice dimension.~\cite{flach2014detangling} We then studied the impact of classical nonlinear interactions on lattices without linear dispersion and formulated necessary and sufficient conditions for nonlinear caging. We used two bands networks as testbeds to show that single particle caging is in general broken by classical interactions. 
We further extended our analysis to three and four band networks, and went into two dimensions with two respectively five band models which again show the possibility of nonlinear caging.

An observation aligned with Ref.~\onlinecite{diliberto2019nonlinear} which follows from the detangling procedure developed in this work is that nonlinear caging is not specific to Kerr nonlinearity but holds also for other, even nonlocal nonlinear terms.
The detangling method leads to a broad number of nontrivial ABF caged model with complicated nonlocal nonlinear interaction terms. Interesting future challenges include the search  for experimentally feasible examples, and their potentially novel features beyond the known ABF models with local Kerr nonlinearity. Another thrilling question we intend to address in a related work concerns the interacting quantum many-body dynamics in ABF networks which satisfy nonlinear caging.

\section{Acknowledgments}

The authors thank Ihor Vakulchyk, Ajith Ramachandran, Arindam Mallick and Tilen \v{C}adez for helpful discussions. This work was supported by the Institute for Basic Science, Korea (IBS-R024-D1).

\appendix

\section{Detangling of ABF one-dimensional networks}
\label{app1}

\subsection{Proof}
\label{app1a}

In this section we provide the proof of the Theorem in Sec.~\ref{sec:abf-gen} stating that any $d=1$ Hamiltonian with all bands flat and finite-range hopping can be recast in a set of decoupled sites via a sequence of local unitary transformation. That is equivalent to show that in these Hamiltonians all compact localized states can be recast to have non-zero amplitude over a single unit cell - showing therefore their orthogonality. Throughout this appendix we will use $\langle A, B\rangle$ to denote the scalar product of matrices: $\langle A, B\rangle = \text{Tr}(A^\dagger B)$ or vectors $\langle A, B\rangle = \sum_a A_a^*B_a$, and we will denote with $u$ the number of unit cells occupied by the CLS.

Let us consider a $d=1$ $\nu$ band Hamiltonian with all bands flat. From Ref.~\onlinecite{read2017compactly} it follows that all of its eigenstates can be represented as spatially compact of size, and we assume that $u\leq \nu$ (in case of flatbands with CLS of different size, we assume that they are all padded by zeros to the size of the largest one). This immediately constrains the possible eigenstates of the Bloch Hamiltonian and it implies that the corresponding Bloch eigenstates take the following form (up to a normalisation prefactor, which still depends on the wavevector $q$):
\begin{gather}
    \Psi_{\alpha,a}(q)  = \frac{\sum_{b=0}^{u-1} C_{\alpha,ab}\, \omega_{qb}}{\sqrt{\langle C_\alpha \omega_q, C_\alpha \omega_q}\rangle} \propto \sum_{b=0}^{u-1} C_{\alpha,ab}\, e^{i b q},\\
    \omega_q = (1, e^{iq}, \dots e^{iq(u-1)}),
\end{gather}
where $\alpha$ is the band index, $a$ is the wave function component. The $\nu\times u$ matrix $C_\alpha$ is parameterising the CLS of band $\alpha$, and the central object of all the following derivations. The eigenstates of a Hermitian Hamiltonian have to be orthogonal, giving the first set of constraints on the matrices $C_\alpha$:
\begin{align}
    \delta_{\alpha\beta} = \sum_a \Psi_{\alpha,a}^*(q)\Psi_{\beta,a}(q) \propto\notag\\
    \label{app:u1p-tab-diagsum-zero}
    \sum_{abc} e^{iq(c-b)} C_{\alpha,ab}^*C_{\beta,ac} = \sum_{bc} e^{iq(c-b)} T_{\alpha\beta,bc}.
\end{align}

We have defined the $\nu\times\nu$ matrices $T_{\alpha\beta} = \langle C_\alpha, C_\beta\rangle$, that will be used later. The above orthogonality condition reduces to a specfic Fourier transform of the matrices $T_{\alpha\beta}$, which implies that the matrices $T_{\alpha\beta}$, $\alpha\neq\beta$ have to have zero sums over any diagonal.

Using this parameterisation of the eigenstates, we can use the spectral decomposition to reconstruct the Hamiltonian itself:
\begin{align}
    \label{eq:Hq_app}
    \mh_q &= M_q\Lambda M_q^\dagger\quad\Lambda_{ab} = \varepsilon_a \delta_{ab},\\
    M_q &= (\Psi_1 \Psi_2 \dots \Psi_\nu), \notag \\
    \Psi_{\alpha,a} &= \frac{C_\alpha \omega_q}{\sqrt{\langle C_\alpha \omega_q, C_\alpha \omega_q\rangle}}.\notag
\end{align}
The Hamiltonian becomes:
\begin{gather*}
    \mh_q = \sum_\alpha \varepsilon_\alpha \frac{(C_\alpha \omega_q)\otimes (C_\alpha \omega_q)^*}{\langle C_\alpha \omega_q, C_\alpha \omega_q\rangle}.
\end{gather*}
Here, $P_\alpha(q) = (C_\alpha \omega_q)\otimes (C_\alpha \omega_q)^*$ and $Q_\alpha(q) = \langle C_\alpha \omega_q, C_\alpha \omega_q\rangle$ are polynomials in $e^{iq}$ of degree $u-1$ and degree at most $u-1$ respectively, and every term in the above sum is their ratio. Therefore the Hamiltonian is long-ranged in general. The above Hamiltonian becomes short-ranged iff $P_\alpha$ is divisible by $Q_\alpha$ $\forall\alpha$. If the degree of $Q_\alpha$ is $u-1$ the ratio $P_\alpha/Q_\alpha$ is a constant, the respective eigenvector is $q$-independent and the CLS is of class $u=1$. Since they are already of class $u=1$, these eigenvalues can be excluded, for example by considering only an orthogonal subspace of the Hilbert space. Therefore I assume that the degree of $Q_\alpha(q)$ is at most $u-2$ in general. This implies that $T_{\alpha\alpha,1u}=T_{\alpha\alpha,u1}=0$ always (see Eq.~\eqref{app:u1p-tab-diagsum-zero}). The lower the degree of $Q_\alpha$ the more zero sum diagonals do $T_{\alpha\alpha}$ have, starting from the corners. Combining this statement with the earlier result that diagonals of $T_{\alpha\beta}$ sum up to zero we see that the amplitudes in the first unit cell of any CLS are always orthogonal to the amplitudes in the last unit cell of any CLS. As we will see below this is the cornerstone of the proof of the triviality of the all bands flat Hamiltonians in $d=1$.

To reconstruct the Hamiltonian we need to find the matrices $C_\alpha$. This requires a solution of a system of coupled matrix quaratic equations with respect to $C_\alpha$, $\alpha=1,\dots,\nu$:
\begin{gather*}
    \langle C_\alpha, C_\beta\rangle = T_{\alpha\beta},
\end{gather*}
considering $T_{\alpha\beta}$ as input parameters. The solution can be constructed sequentially: we parameterise $C_\alpha = (c_{\alpha1}, c_{\alpha2},\dots c_{\alpha u})$, where $c_{\alpha a}$ is a vector of the eigenfunction amplitudes in the unit cell $a$ of the CLS of the band $\alpha$. This transforms the above equations into a set of coupled quadratic equations for $c_{\alpha a}$. We solve these equations iteratively by fixing $c_{\alpha,a}$ one by one starting from $\alpha=1$ and only taking into account the equations involving $\beta\leq\alpha$. We also employ extensively our freedom in the choice of the basis vector of the Hilbert space, to simplify the solution. The core idea is to see how the equations constrain the possible shapes of $c_{\alpha,1}$ and $c_{\alpha,\nu}$ and show that one can always redefine the unit cell to reduce the sizes of all the CLS by $1$.

We start by setting $c_{11} = e_1$ - this defines the first basis vector. Then $T_{11,1u}=0$ implies that we can define $c_{1u} = e_2$. For $c_{21}$ we have the following constraint:
\begin{gather*}
    T_{12,u 1} = \langle c_{1u}, c_{21}\rangle = 0,
\end{gather*}
and we can choose $c_{21}=*e_1 + *e_3$, where we defined the next basis vector $e_3$ and the asterisks stand for some (possibly zero) coefficients. The $c_{2u}$ is constrained by
\begin{gather*}
    T_{21,1u} = \langle c_{11}, c_{2u}\rangle = 0,\\
    T_{22,1u} = \langle c_{21}, c_{2u}\rangle = 0.
\end{gather*}
The most generic form of $c_{2u} = *e_2 + *e_4$ - again defining the basis vector $e_4$. Such incremental construction enforces $c_{\beta u}$ to have zeros at positions where $c_{\alpha1}$ has non-zero elements and vice versa. It therefore guarantees the existence of a pattern of non-zero elements in $C_\alpha$ that is the same $\forall\,\alpha$.

We illustrate this result by a specific case of $\nu=3$, $u=3$ (there is a single possible redefinition of the unit cell in the case of $\nu=2$ and $u=2$, that we discuss in the next appendix): working out the matrices $C_1$, $C_2$, $C_3$ following the above rules we find:
\begin{gather*}
    C_1 = \begin{pmatrix}
        * & * & 0\\
        0 & * & *\\
        0 & * & 0 \\
    \end{pmatrix},
    C_2 = \begin{pmatrix}
        * & * & 0\\
        0 & * & *\\
        * & * & 0 \\
    \end{pmatrix},
    C_3 = \begin{pmatrix}
        * & * & 0\\
        0 & * & *\\
        * & * & 0 \\
    \end{pmatrix}.
\end{gather*}
As before the asterisks $*$ denote unspecified coefficients. The shape of the all the matrices supports a redefinition of the unit cell, that reduces the sizes of all the CLS/matrices to $u=2$.

\subsection{Parametrization of $\nu=2$ networks}
\label{app1b}

\begin{figure}[!htbp]
    \centering
    \includegraphics [width=0.7\columnwidth]{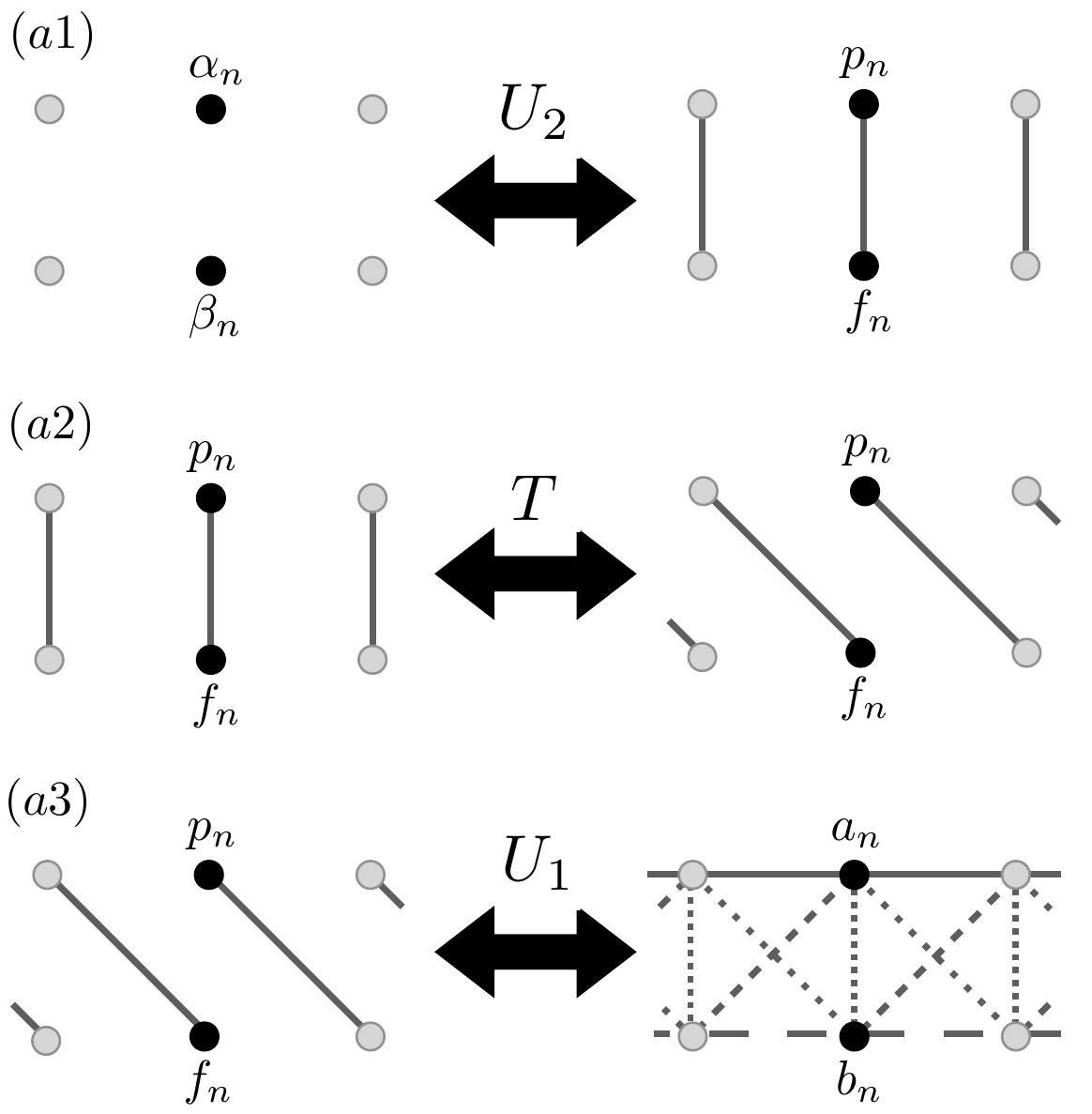}
    \caption{Schematic representation of the unit cell redefinition for $\nu=2$ bands lattice. In each panel, the black dots represent the chosen unit cell.}
    \label{fig:UC_redef_app}
\end{figure}

The above proof when inverted yields a generator scheme for dispersionless networks. In this subsection, we explicitly unfold the two bands problem, $\nu=2$. Let us consider a non-degenerate (two different flatband energies) fully decoupled network in coordinates $\phi_n=(\alpha_n,\beta_n)$.
\begin{gather}
    \label{eq:fb_eq_app}
    i \dot{\phi}_n = - H_0^{(1)}\phi_n -  H_1^{(1)}\phi_{n+1} - H_1^{(1)\dagger}\phi_{n-1},\\
    H_0^{(1)} =
    \begin{pmatrix}
        -1 &  0  \\[0.3em]
        0 & 1
    \end{pmatrix}
    \qquad
    H_1^{(1)} =
    \begin{pmatrix}
        0 &  0  \\[0.3em]
        0 & 0
    \end{pmatrix},
    \label{eq:H0H1_app}
\end{gather}
with two flatband energies $E_1 = -1$ and $E_2=1$. 
The mapping in Eq.~\eqref{eq:par_nu2} consists of two unitary transformations $U_1$ and $U_2$, which are parametrized as
\begin{gather}
    U_i = e^{i\theta_i}
    \begin{pmatrix}
        z_i &  w_i   \\[0.3em]
        -  w^*_i & z^*_i
    \end{pmatrix}
    \qquad i=1,2
    \label{eq:U12_app}
\end{gather}
by the complex numbers $z_i,w_i$ such that $|z_i|^2 + |w_i|^2 = 1$ and the phases $\theta_i$. The unit cell redefinition $T$ is
\begin{gather}
    T:\ 
    \begin{cases}
        p_n \longmapsto p_n\\
        f_n \longmapsto f_{n-1}
    \end{cases}
    \label{eq:UC_redef_app}
\end{gather}
We generate all the $\nu=2$ dispersionless lattices by applying the transformations, Eqs.~(\ref{eq:U12_app},\ref{eq:UC_redef_app}) in the following order
\begin{gather}
    \label{eq:par_nu2_app}
    \begin{pmatrix}
        \alpha_n  \\[0.3em]
        \beta_n 
    \end{pmatrix}
    \xmapsto{\ \ U_2\ \ }
    \begin{pmatrix}
        p_n  \\[0.3em]
        f_n 
    \end{pmatrix}
    \xmapsto{\ \ T\ \ }
    \begin{pmatrix}
        p_n  \\[0.3em]
        f_n 
    \end{pmatrix}
    \xmapsto{\ \  U_1\ \  }
    \begin{pmatrix}
        a_n  \\[0.3em]
        b_n 
    \end{pmatrix},
\end{gather}
as shown in Fig.~\ref{fig:UC_redef_app}.

\noindent
The first coordinate rotation $U_2$ turns $H_0^{(1)}$ in Eq.~\eqref{eq:H0H1_app} to 
\begin{align}
    H_0^{(2)} &= U_2 H_0^{(1)} U_2^\dagger =
    \begin{pmatrix}
        |w_2|^2  -  |z_2|^2 &  2  z_2 w_2 \\[0.3em]
        2 z_2^* w_2^* & |z_2|^2  -  |w_2|^2
    \end{pmatrix},
    \label{eq:H0H1_2_rot_app}
\end{align} 
while $H_1^{(2)} = U_2 H_1^{(1)} U_2^\dagger$ remains zero - as shown in Fig.~\ref{fig:UC_redef_app}(a1), right plot. 

\noindent The unit cell redefinition $T$ in Eq.~\ref{eq:UC_redef_app}) redefines $H_0^{(2)}$ and $H_1^{(2)}$ as 
\begin{align}
    \label{eq:H0H1_3_rot0_app}
    H_0^{(3)} &= 
    \begin{pmatrix}
        |w_2|^2  -  |z_2|^2 &  0 \\[0.3em]
        0 & |z_2|^2  -  |w_2|^2
    \end{pmatrix},\\
    &H_1^{(3)}=
    \begin{pmatrix}
        0 &  2  z_2 w_2   \\[0.3em]
        0 &0 
    \end{pmatrix},
    \label{eq:H0H1_3_rot1_app}
\end{align}
as shown in Fig.~\ref{fig:UC_redef_app}(a2), right plot.

\noindent At last, the rotation $U_1$ turns $H_0^{(3)}$ and $H_1^{(3)}$ in Eqs.~(\ref{eq:H0H1_3_rot0_app},\ref{eq:H0H1_3_rot1_app}) to 
the following matrices 
\begin{align}
    \label{eq:H0H1_rot0_app}
    H_0 &= U_1 H_0^{(3)} U_1^\dagger = \Gamma_0
    \begin{pmatrix}
        |z_1|^2  -  |w_1|^2 &  -2  z_1 w_1 \\[0.3em]
        -2 z_1^* w_1^* &   |w_1|^2 - |z_1|^2
    \end{pmatrix},\\
    &H_1 = U_1 H_1^{(3)} U_1^\dagger = \Gamma_1
    \begin{pmatrix}
        z_1  w_1^* &  z_1^2  \\[0.3em]
        - (w_1^*)^2 &- z_1  w_1^*
    \end{pmatrix},
    \label{eq:H0H1_rot1_app}
\end{align}
for $\Gamma_0 =  |w_2|^2 -  |z_2|^2$ and $\Gamma_1 = 2 z_2 w_2$ - as shown in Fig.~\ref{fig:UC_redef_app}(a3), right plot.

\section{Detangling procedure applied to nonlinear dispersionless models}
\label{app2}

In this appendix, we apply the detangling procedure as described by Eq.~\eqref{eq:par_nu2} to the dispersionless, two band models in presence of a local Kerr nonlinearity~\eqref{eq:dnls_ham_NL}.

\subsection{Preserving the caging in $\nu=2$ networks: a necessary and sufficient condition}
\label{app2c}

We now work out the detangled local Kerr nonlinearity for a general $\nu=2$ AB lattice. The transformation $U_1$ in components read
\begin{gather}
    U_1\ :\ 
    \begin{cases}
        a_n = e^{i\theta_1}(z_1 p_n + w_1 f_n) \\
        b_n = e^{i\theta_1}(-w_1^* p_n + z_1^* f_n) \\
    \end{cases}
    \label{eq:U1_app}
\end{gather}
Via Eq.~\eqref{eq:U1_app}, the nonlinear terms turn
\begin{align}
    \hspace{-5mm}
    a_n|a_n|^2  =  e^{i\theta_1}&\left[ z_1 |z_1|^2 p_n|p_n|^2 + z_1^2 w_1^* p_n^2 f_n^* \right.\notag\\
    &\left.+ z_1^* w_1^2  p_n^* f_n^2 + w_1 |w_1|^2 f_n|f_n|^2\right. \notag \\
    & \left.+ 2  |z_1|^2 w_1 |p_n|^2 f_n + 2 z_1 |w_1|^2p_n |f_n|^2\right], \notag \\
    \vspace{2mm}
    b_n|b_n|^2 
    \label{eq:nonlin_terms1_app}
    =e^{i\theta_1}&\left[ - w_1^* |w_1|^2 p_n|p_n|^2 + z_1 w_1^{*2} p_n^2 f_n^* \right.\\
    &\left. - z_1^{*2} w_1 p_n^*  f_n^2 +  z_1^* |z_1|^2  f_n|f_n|^2 \right. \notag \\
    &\left.+  2 z_1^* |w_1|^2 |p_n|^2 f_n  -  2 |z_1|^2 w_1^* p_n |f_n|^2 \right]. \notag
\end{align}
The equations for $p_n$ then read
\begin{align}
    i\dot{p}_n &= -(|z_2|^2  -  |w_2|^2) p_n + 2 w_2 z_2 f_{n-1} \notag \\
    & +U\left\{ \left(|z_1|^4 + |w_1|^4 \right) p_n|p_n|^2  \right. \notag \\
    &\qquad\quad\left. +  2 z_1^{*2} w_1^2\, p_n^* f_n^2  +   4 |z_1|^2 |w_1|^2 \, p_n |f_n|^2 \right. \notag\\
    \label{eq:nu2_pn_app}
    &\qquad\quad \left.  + z_1 w_1^* \left( |z_1|^2 - |w_1|^2  \right) p_n^2 f_n^* \right. \\
    &\qquad\quad \left.  +  z_1^* w_1 \left( |w_1|^2 - |z_1|^2  \right)    f_n|f_n|^2     \right. \notag \\
    &\qquad\quad \left.  + 2 z_1^* w_1 \left( |z_1|^2 - |w_1|^2  \right)      |p_n|^2 f_n
    \right\}. \notag
\end{align}
The unit cell redefinition $f_n\longmapsto f_{n-1}$ yields the terms $f_{n-1}|f_{n-1}|^2$ 
\begin{align}
    i\dot{p}_n &= -(|z_2|^2  -  |w_2|^2) p_n + 2 w_2 z_2 f_{n} \notag \\
    & +U\left\{ \left(|z_1|^4 + |w_1|^4 \right)  p_n|p_n|^2  \right. \notag \\
    &\qquad\quad \left.  +  2 z_1^{*2} w_1^2 \, p_n^* f_{n-1}^2 + 4 |z_1|^2 |w_1|^2 \, p_n |f_{n-1}|^2\right. \notag\\
    \label{eq:nu2_pn2_app}
    &\qquad\quad \left.  + z_1w_1^* \left( |z_1|^2 - |w_1|^2  \right) p_n^2 f_{n-1}^* \right. \\
    &\qquad\quad \left.  +  z_1^* w_1 \left( |w_1|^2 - |z_1|^2  \right)    f_{n-1}|f_{n-1}|^2     \right. \notag \\
    &\qquad\quad \left.  + 2 z_1^* w_1 \left( |z_1|^2 - |w_1|^2  \right)      |p_n|^2 f_{n-1}   
    \right\} \notag
\end{align}
which break the caging effect. These terms are not present if $|w_1|^2 = |z_1|^2$, which reduces Eq.~\eqref{eq:nu2_pn2_app} to 
\begin{align}
    \label{eq:nu2_pn3_app}
    i\dot{p}_n &= -(|z_2|^2  -  |w_2|^2) p_n + 2 w_2 z_2 f_{n} \\
    & + 2 e^{i\theta_1} U\left\{ |z_1|^4 (p_n|p_n|^2 + 2 p_n |f_{n-1}|^2)  +   z_1^{*2} w_1^2 \, p_n^* f_{n-1}^2 
    \right\}.\notag
\end{align} 
Similarly follows for $f_n$ where under the condition $|w_1|^2 = |z_1|^2$ the motion equation reads 
\begin{align}
    \label{eq:nu2_fn2_app}
    i\dot{f}_n &= (|z_2|^2  -  |w_2|^2) f_n + 2 w_2^* z_2^* p_{n} \\
    &+ 2  U\left\{ |z_1|^4   (f_n|f_n|^2 + 2|p_{n+1}|^2  f_n )  +   z_1^{2} w_1^{*2}     p_{n+1}^2 f_n^*   
    \right\}. \notag
\end{align}

\noindent
As a result we observe the following:
\begin{enumerate}
    \item the condition $|w_1|^2 = |z_1|^2$ in Eq.~\eqref{eq:U12_app} is \emph{necessary}  to preserve the caging  - otherwise "fully nonlocal" terms which break the caging exist in Eq.~\ref{eq:nu2_pn2_app} - and it is \emph{sufficient} - since the subsequent transformation by $U_2$ in Eq.~\eqref{eq:U12_app} will not introduce additional "fully nonlocal" terms.
    \item the condition $|w_1|^2 = |z_1|^2$ in Eq.~\eqref{eq:U12_app} yields all the entrees of the matrix $H_1$ in Eq.~\eqref{eq:H0H1_rot1_app} have equal magnitude in absolute value. 
\end{enumerate}

\subsection{Rotating the interaction Hamiltonian $\mh_1^G$}
\label{app3a}

The rotated Eqs.~(\ref{eq:nu2_pn3_app},\ref{eq:nu2_fn2_app})  are the equations of motion $i\dot{p}_n=\frac{\partial \mh_G}{\partial p_n^{\ast}}$ and $i\dot{f}_n=\frac{\partial \mh}{\partial f_n^{\ast}}$ of the Hamiltonian $\mh_G$
\begin{gather}
    \mh_G = \mh_0^G + \mh_1^G,
    \label{eq:Ham_app}
\end{gather}
where 
\begin{align}
    \mh_0^G&=  \sum_n \left\{ (|z_2|^2  -  |w_2|^2) (|f_n|^2 - |p_n|^2) \right. \notag \\
    &\left. \qquad\quad  +2 w_2 z_2 p_n^* f_n + H.c. \right\},
    \label{eq:Ham0_app}
\end{align}
and
\begin{align}
    \mh_1^G &= U\sum_n  \left\{  |z_1|^4  \left[ |p_n|^4 + |f_n|^4 + 4 |p_{n}|^2 |f_{n+1}|^2 \right] \right. \notag \\
    &\left. \qquad\quad  + z_1^{*2} w_1^2 p_n^{*2} f_{n+1}^2 + z_1 w_1^{*2} p_n^{2} f_{n+1}^{*2} \right\}.
    \label{eq:Ham1_app}
\end{align}

\subsection{Fully detangled models A and B}
\label{app:fd-ab}

By applying further transformations (see Fig.~\ref{fig:abf_rot}) one can fully detangle both models A and B introduced in the main text. This computation is straightforward but lengthy and we only provide the final result for both models. Namely the nonlinear evolution equation for model A is
\begin{align}
    i\dot{\alpha}_n = 2\alpha_n  + &\frac{U}{8}  \left\{\alpha_n|\alpha_n|^2 + \beta_n^2 \alpha_n^*  + 2\alpha_n |\beta_n|^2 \right. \notag \\
    & \left. + \alpha_n^*\alpha_{n+1}^2 + \alpha_n^*\beta_{n+1}^2 - 2\alpha_n^*\alpha_{n+1} \beta_{n+1}   \right. \notag \\
    & \left. +   \beta_n^*\alpha_{n+1}^2 +  \beta_n^*\beta_{n+1}^2 - 2 \beta_n^*\alpha_{n+1} \beta_{n+1} \right. \notag \\
    &\left.+ \alpha_n^*\alpha_{n-1}^2 + \alpha_n^*\beta_{n-1}^2 + 2\alpha_n^*\alpha_{n-1} \beta_{n-1}   \right. \notag \\
    & \left.   -  \beta_n^*\alpha_{n-1}^2 -  \beta_n^*\beta_{n-1}^2 - 2 \beta_n^*\alpha_{n-1} \beta_{n-1}  \right. \notag \\
    &\left. + 2 ( \alpha_n |\alpha_{n+1}|^2 - \alpha_n\alpha_{n+1}^*\beta_{n+1}   \right. \\
    & \left. \qquad - \alpha_n\alpha_{n+1}\beta_{n+1}^* + \alpha_n|\beta_{n+1}|^2) \right. \notag \\
    &\left. +2( \beta_n |\alpha_{n+1}|^2 - \alpha_{n+1}^* \beta_n \beta_{n+1}  \right. \notag \\
    & \left. \qquad  - \alpha_{n+1} \beta_n \beta_{n+1}^* + \beta_n|\beta_{n+1}|^2) \right. \notag \\
    &\left. + 2 ( \alpha_n |\alpha_{n-1}|^2 + \alpha_n\alpha_{n-1}^*\beta_{n-1} \right. \notag \\
    & \left. \qquad   + \alpha_n\alpha_{n-1}\beta_{n-1}^* + \alpha_n|\beta_{n-1}|^2) \right. \notag \\
    &\left. -2( \beta_n |\alpha_{n-1}|^2 + \alpha_{n-1}^* \beta_n \beta_{n-1} \right. \notag \\
    & \left. \qquad  + \alpha_{n-1} \beta_n \beta_{n-1}^* + \beta_n|\beta_{n+1}|^2)\right\} \notag
\end{align}
and a similar equation for the other component $\beta_n$. It is straightforward to check that all the terms in the above expression are exactly of the type required by the nonlinear caging criterion~\eqref{eq:gen-ue-caged} introduced in Sec.~\ref{sec:iff-caging}.

The equation for model B reads (again we only provide the equation for one component, the equation for the other component $\beta_n$ is qualitatively similar):
\begin{align}
    i\dot{\alpha}_n = 4\alpha_n  + &\frac{U}{64}  \left\{ \right. 2 \sqrt{3} \left[\alpha_{n+1}|\alpha_{n+1}|^2   - \alpha_{n-1}|\alpha_{n-1}|^2   \right. \notag \\
    &\left. - \alpha_{n+1}^2 \beta_{n+1}^* - \alpha_{n-1}^2 \beta_{n-1}^* \right. \notag \\
    &\left. + \beta_{n+1}^2 \alpha_{n+1}^* - \beta_{n-1}^2 \alpha_{n-1}^* \right. \\
    &\left. -  \beta_{n+1}|\beta_{n+1}|^2 -  \beta_{n-1}|\beta_{n-1}|^2  \right. \notag \\
    &\left. - 2 |\alpha_{n+1}|^2 \beta_{n+1} - 2 |\alpha_{n-1}|^2 \beta_{n-1} \right. \notag \\
    &\left. + 2\alpha_{n+1} |\beta_{n+1}|^2 - 2\alpha_{n-1} |\beta_{n-1}|^2 \right. \notag \\
    &\left. + \left[\dots \right]  \right\}, \notag
\end{align}
We have only kept the terms in the above expression, that break the nonlinear caging criterion~\eqref{eq:gen-ue-caged}. The remaining terms, indicated by $ \left[\dots \right]$ conform with the caging criterion.

\bibliography{flatband,mbl,general}

\begin{thebibliography}{49}%
\makeatletter
\providecommand \@ifxundefined [1]{%
 \@ifx{#1\undefined}
}%
\providecommand \@ifnum [1]{%
 \ifnum #1\expandafter \@firstoftwo
 \else \expandafter \@secondoftwo
 \fi
}%
\providecommand \@ifx [1]{%
 \ifx #1\expandafter \@firstoftwo
 \else \expandafter \@secondoftwo
 \fi
}%
\providecommand \natexlab [1]{#1}%
\providecommand \enquote  [1]{``#1''}%
\providecommand \bibnamefont  [1]{#1}%
\providecommand \bibfnamefont [1]{#1}%
\providecommand \citenamefont [1]{#1}%
\providecommand \href@noop [0]{\@secondoftwo}%
\providecommand \href [0]{\begingroup \@sanitize@url \@href}%
\providecommand \@href[1]{\@@startlink{#1}\@@href}%
\providecommand \@@href[1]{\endgroup#1\@@endlink}%
\providecommand \@sanitize@url [0]{\catcode `\\12\catcode `\$12\catcode
  `\&12\catcode `\#12\catcode `\^12\catcode `\_12\catcode `\%12\relax}%
\providecommand \@@startlink[1]{}%
\providecommand \@@endlink[0]{}%
\providecommand \url  [0]{\begingroup\@sanitize@url \@url }%
\providecommand \@url [1]{\endgroup\@href {#1}{\urlprefix }}%
\providecommand \urlprefix  [0]{URL }%
\providecommand \Eprint [0]{\href }%
\providecommand \doibase [0]{http://dx.doi.org/}%
\providecommand \selectlanguage [0]{\@gobble}%
\providecommand \bibinfo  [0]{\@secondoftwo}%
\providecommand \bibfield  [0]{\@secondoftwo}%
\providecommand \translation [1]{[#1]}%
\providecommand \BibitemOpen [0]{}%
\providecommand \bibitemStop [0]{}%
\providecommand \bibitemNoStop [0]{.\EOS\space}%
\providecommand \EOS [0]{\spacefactor3000\relax}%
\providecommand \BibitemShut  [1]{\csname bibitem#1\endcsname}%
\let\auto@bib@innerbib\@empty
\bibitem [{\citenamefont {Anderson}(1958)}]{anderson1958absence}%
  \BibitemOpen
  \bibfield  {author} {\bibinfo {author} {\bibfnamefont {P.~W.}\ \bibnamefont
  {Anderson}},\ }\bibfield  {title} {\enquote {\bibinfo {title} {Absence of
  diffusion in certain random lattices},}\ }\href {\doibase
  10.1103/PhysRev.109.1492} {\bibfield  {journal} {\bibinfo  {journal} {Phys.
  Rev.}\ }\textbf {\bibinfo {volume} {109}},\ \bibinfo {pages} {1492--1505}
  (\bibinfo {year} {1958})}\BibitemShut {NoStop}%
\bibitem [{\citenamefont {Kramer}\ and\ \citenamefont
  {MacKinnon}(1993)}]{kramer1993localization}%
  \BibitemOpen
  \bibfield  {author} {\bibinfo {author} {\bibfnamefont {B.}~\bibnamefont
  {Kramer}}\ and\ \bibinfo {author} {\bibfnamefont {A.}~\bibnamefont
  {MacKinnon}},\ }\bibfield  {title} {\enquote {\bibinfo {title} {Localization:
  theory and experiment},}\ }\href {\doibase 10.1088/0034-4885/56/12/001}
  {\bibfield  {journal} {\bibinfo  {journal} {Rep. Prog. Phys.}\ }\textbf
  {\bibinfo {volume} {56}},\ \bibinfo {pages} {1469--1564} (\bibinfo {year}
  {1993})}\BibitemShut {NoStop}%
\bibitem [{\citenamefont {Basko}\ \emph {et~al.}(2006)\citenamefont {Basko},
  \citenamefont {Aleiner},\ and\ \citenamefont {Altshuler}}]{basko2006metal}%
  \BibitemOpen
  \bibfield  {author} {\bibinfo {author} {\bibfnamefont {D.M.}\ \bibnamefont
  {Basko}}, \bibinfo {author} {\bibfnamefont {I.L.}\ \bibnamefont {Aleiner}}, \
  and\ \bibinfo {author} {\bibfnamefont {B.L.}\ \bibnamefont {Altshuler}},\
  }\bibfield  {title} {\enquote {\bibinfo {title} {Metal–insulator transition
  in a weakly interacting many-electron system with localized single-particle
  states},}\ }\href {\doibase 10.1016/j.aop.2005.11.014} {\bibfield  {journal}
  {\bibinfo  {journal} {Ann. Phys.}\ }\textbf {\bibinfo {volume} {321}},\
  \bibinfo {pages} {1126 -- 1205} (\bibinfo {year} {2006})}\BibitemShut
  {NoStop}%
\bibitem [{\citenamefont {Aleiner}\ \emph {et~al.}(2010)\citenamefont
  {Aleiner}, \citenamefont {Altshuler},\ and\ \citenamefont
  {Shlyapnikov}}]{aleiner2010finite}%
  \BibitemOpen
  \bibfield  {author} {\bibinfo {author} {\bibfnamefont {I.~L.}\ \bibnamefont
  {Aleiner}}, \bibinfo {author} {\bibfnamefont {B.~L.}\ \bibnamefont
  {Altshuler}}, \ and\ \bibinfo {author} {\bibfnamefont {G.~V.}\ \bibnamefont
  {Shlyapnikov}},\ }\bibfield  {title} {\enquote {\bibinfo {title} {A
  finite-temperature phase transition for disordered weakly interacting bosons
  in one dimension},}\ }\href {\doibase 10.1038/nphys1758} {\bibfield
  {journal} {\bibinfo  {journal} {Nat. Phys.}\ }\textbf {\bibinfo {volume}
  {6}},\ \bibinfo {pages} {900--904} (\bibinfo {year} {2010})}\BibitemShut
  {NoStop}%
\bibitem [{\citenamefont {Abanin}\ \emph {et~al.}(2019)\citenamefont {Abanin},
  \citenamefont {Altman}, \citenamefont {Bloch},\ and\ \citenamefont
  {Serbyn}}]{abanin2019colloquium}%
  \BibitemOpen
  \bibfield  {author} {\bibinfo {author} {\bibfnamefont {Dmitry~A.}\
  \bibnamefont {Abanin}}, \bibinfo {author} {\bibfnamefont {Ehud}\ \bibnamefont
  {Altman}}, \bibinfo {author} {\bibfnamefont {Immanuel}\ \bibnamefont
  {Bloch}}, \ and\ \bibinfo {author} {\bibfnamefont {Maksym}\ \bibnamefont
  {Serbyn}},\ }\bibfield  {title} {\enquote {\bibinfo {title} {Colloquium:
  Many-body localization, thermalization, and entanglement},}\ }\href {\doibase
  10.1103/RevModPhys.91.021001} {\bibfield  {journal} {\bibinfo  {journal}
  {Rev. Mod. Phys.}\ }\textbf {\bibinfo {volume} {91}},\ \bibinfo {pages}
  {021001} (\bibinfo {year} {2019})}\BibitemShut {NoStop}%
\bibitem [{\citenamefont {Laptyeva}\ \emph {et~al.}(2014)\citenamefont
  {Laptyeva}, \citenamefont {Ivanchenko},\ and\ \citenamefont
  {Flach}}]{laptyeva2014nonlinear}%
  \BibitemOpen
  \bibfield  {author} {\bibinfo {author} {\bibfnamefont {T~V}\ \bibnamefont
  {Laptyeva}}, \bibinfo {author} {\bibfnamefont {M~V}\ \bibnamefont
  {Ivanchenko}}, \ and\ \bibinfo {author} {\bibfnamefont {S}~\bibnamefont
  {Flach}},\ }\bibfield  {title} {\enquote {\bibinfo {title} {Nonlinear lattice
  waves in heterogeneous media},}\ }\href {\doibase
  10.1088/1751-8113/47/49/493001} {\bibfield  {journal} {\bibinfo  {journal} {J
  Phys. A: Math. Theor.}\ }\textbf {\bibinfo {volume} {47}},\ \bibinfo {pages}
  {493001} (\bibinfo {year} {2014})}\BibitemShut {NoStop}%
\bibitem [{\citenamefont {Vakulchyk}\ \emph {et~al.}(2019)\citenamefont
  {Vakulchyk}, \citenamefont {Fistul},\ and\ \citenamefont
  {Flach}}]{vakulchyk2019wave}%
  \BibitemOpen
  \bibfield  {author} {\bibinfo {author} {\bibfnamefont {Ihor}\ \bibnamefont
  {Vakulchyk}}, \bibinfo {author} {\bibfnamefont {Mikhail~V.}\ \bibnamefont
  {Fistul}}, \ and\ \bibinfo {author} {\bibfnamefont {Sergej}\ \bibnamefont
  {Flach}},\ }\bibfield  {title} {\enquote {\bibinfo {title} {Wave packet
  spreading with disordered nonlinear discrete-time quantum walks},}\ }\href
  {\doibase 10.1103/PhysRevLett.122.040501} {\bibfield  {journal} {\bibinfo
  {journal} {Phys. Rev. Lett.}\ }\textbf {\bibinfo {volume} {122}},\ \bibinfo
  {pages} {040501} (\bibinfo {year} {2019})}\BibitemShut {NoStop}%
\bibitem [{\citenamefont {Mielke}(1991)}]{mielke1991ferromagnetic}%
  \BibitemOpen
  \bibfield  {author} {\bibinfo {author} {\bibfnamefont {A}~\bibnamefont
  {Mielke}},\ }\bibfield  {title} {\enquote {\bibinfo {title} {Ferromagnetic
  ground states for the hubbard model on line graphs},}\ }\href
  {http://stacks.iop.org/0305-4470/24/i=2/a=005} {\bibfield  {journal}
  {\bibinfo  {journal} {J Phys. A: Math. and Gen.}\ }\textbf {\bibinfo {volume}
  {24}},\ \bibinfo {pages} {L73} (\bibinfo {year} {1991})}\BibitemShut
  {NoStop}%
\bibitem [{\citenamefont {Tasaki}(1992)}]{tasaki1992ferromagnetism}%
  \BibitemOpen
  \bibfield  {author} {\bibinfo {author} {\bibfnamefont {Hal}\ \bibnamefont
  {Tasaki}},\ }\bibfield  {title} {\enquote {\bibinfo {title} {Ferromagnetism
  in the hubbard models with degenerate single-electron ground states},}\
  }\href {\doibase 10.1103/PhysRevLett.69.1608} {\bibfield  {journal} {\bibinfo
   {journal} {Phys. Rev. Lett.}\ }\textbf {\bibinfo {volume} {69}},\ \bibinfo
  {pages} {1608--1611} (\bibinfo {year} {1992})}\BibitemShut {NoStop}%
\bibitem [{\citenamefont {Derzhko}\ \emph {et~al.}(2015)\citenamefont
  {Derzhko}, \citenamefont {Richter},\ and\ \citenamefont
  {Maksymenko}}]{derzhko2015strongly}%
  \BibitemOpen
  \bibfield  {author} {\bibinfo {author} {\bibfnamefont {Oleg}\ \bibnamefont
  {Derzhko}}, \bibinfo {author} {\bibfnamefont {Johannes}\ \bibnamefont
  {Richter}}, \ and\ \bibinfo {author} {\bibfnamefont {Mykola}\ \bibnamefont
  {Maksymenko}},\ }\bibfield  {title} {\enquote {\bibinfo {title} {Strongly
  correlated flat-band systems: The route from heisenberg spins to hubbard
  electrons},}\ }\href {\doibase 10.1142/S0217979215300078} {\bibfield
  {journal} {\bibinfo  {journal} {Int. J. Mod. Phys. B}\ }\textbf {\bibinfo
  {volume} {29}},\ \bibinfo {pages} {1530007} (\bibinfo {year}
  {2015})}\BibitemShut {NoStop}%
\bibitem [{\citenamefont {Leykam}\ \emph {et~al.}(2018)\citenamefont {Leykam},
  \citenamefont {Andreanov},\ and\ \citenamefont
  {Flach}}]{leykam2018artificial}%
  \BibitemOpen
  \bibfield  {author} {\bibinfo {author} {\bibfnamefont {Daniel}\ \bibnamefont
  {Leykam}}, \bibinfo {author} {\bibfnamefont {Alexei}\ \bibnamefont
  {Andreanov}}, \ and\ \bibinfo {author} {\bibfnamefont {Sergej}\ \bibnamefont
  {Flach}},\ }\bibfield  {title} {\enquote {\bibinfo {title} {Artificial flat
  band systems: from lattice models to experiments},}\ }\href {\doibase
  10.1080/23746149.2018.1473052} {\bibfield  {journal} {\bibinfo  {journal}
  {Adv. Phys.: X}\ }\textbf {\bibinfo {volume} {3}},\ \bibinfo {pages}
  {1473052} (\bibinfo {year} {2018})}\BibitemShut {NoStop}%
\bibitem [{\citenamefont {Leykam}\ and\ \citenamefont
  {Flach}(2018)}]{leykam2018perspective}%
  \BibitemOpen
  \bibfield  {author} {\bibinfo {author} {\bibfnamefont {Daniel}\ \bibnamefont
  {Leykam}}\ and\ \bibinfo {author} {\bibfnamefont {Sergej}\ \bibnamefont
  {Flach}},\ }\bibfield  {title} {\enquote {\bibinfo {title} {Perspective:
  Photonic flatbands},}\ }\href {\doibase 10.1063/1.5034365} {\bibfield
  {journal} {\bibinfo  {journal} {APL Phot.}\ }\textbf {\bibinfo {volume}
  {3}},\ \bibinfo {pages} {070901} (\bibinfo {year} {2018})}\BibitemShut
  {NoStop}%
\bibitem [{\citenamefont {Vidal}\ \emph {et~al.}(1998)\citenamefont {Vidal},
  \citenamefont {Mosseri},\ and\ \citenamefont {Dou\ifmmode~\mbox{\c{c}}\else
  \c{c}\fi{}ot}}]{vidal1998aharonov}%
  \BibitemOpen
  \bibfield  {author} {\bibinfo {author} {\bibfnamefont {Julien}\ \bibnamefont
  {Vidal}}, \bibinfo {author} {\bibfnamefont {R\'emy}\ \bibnamefont {Mosseri}},
  \ and\ \bibinfo {author} {\bibfnamefont {Benoit}\ \bibnamefont
  {Dou\ifmmode~\mbox{\c{c}}\else \c{c}\fi{}ot}},\ }\bibfield  {title} {\enquote
  {\bibinfo {title} {{Aharonov-Bohm} cages in two-dimensional structures},}\
  }\href {\doibase 10.1103/PhysRevLett.81.5888} {\bibfield  {journal} {\bibinfo
   {journal} {Phys. Rev. Lett.}\ }\textbf {\bibinfo {volume} {81}},\ \bibinfo
  {pages} {5888--5891} (\bibinfo {year} {1998})}\BibitemShut {NoStop}%
\bibitem [{\citenamefont {Dou\ifmmode~\mbox{\c{c}}\else \c{c}\fi{}ot}\ and\
  \citenamefont {Vidal}(2002)}]{doucot2002pairing}%
  \BibitemOpen
  \bibfield  {author} {\bibinfo {author} {\bibfnamefont {Benoit}\ \bibnamefont
  {Dou\ifmmode~\mbox{\c{c}}\else \c{c}\fi{}ot}}\ and\ \bibinfo {author}
  {\bibfnamefont {Julien}\ \bibnamefont {Vidal}},\ }\bibfield  {title}
  {\enquote {\bibinfo {title} {Pairing of cooper pairs in a fully frustrated
  josephson-junction chain},}\ }\href {\doibase 10.1103/PhysRevLett.88.227005}
  {\bibfield  {journal} {\bibinfo  {journal} {Phys. Rev. Lett.}\ }\textbf
  {\bibinfo {volume} {88}},\ \bibinfo {pages} {227005} (\bibinfo {year}
  {2002})}\BibitemShut {NoStop}%
\bibitem [{\citenamefont {Fang}\ \emph
  {et~al.}(2012{\natexlab{a}})\citenamefont {Fang}, \citenamefont {Yu},\ and\
  \citenamefont {Fan}}]{fang2012photonic}%
  \BibitemOpen
  \bibfield  {author} {\bibinfo {author} {\bibfnamefont {Kejie}\ \bibnamefont
  {Fang}}, \bibinfo {author} {\bibfnamefont {Zongfu}\ \bibnamefont {Yu}}, \
  and\ \bibinfo {author} {\bibfnamefont {Shanhui}\ \bibnamefont {Fan}},\
  }\bibfield  {title} {\enquote {\bibinfo {title} {{Photonic Aharonov-Bohm
  Effect Based on Dynamic Modulation}},}\ }\href {\doibase
  10.1103/PhysRevLett.108.153901} {\bibfield  {journal} {\bibinfo  {journal}
  {Phys. Rev. Lett.}\ }\textbf {\bibinfo {volume} {108}},\ \bibinfo {pages}
  {153901} (\bibinfo {year} {2012}{\natexlab{a}})}\BibitemShut {NoStop}%
\bibitem [{\citenamefont {Longhi}(2014)}]{longhi2014aharonov}%
  \BibitemOpen
  \bibfield  {author} {\bibinfo {author} {\bibfnamefont {Stefano}\ \bibnamefont
  {Longhi}},\ }\bibfield  {title} {\enquote {\bibinfo {title} {{Aharonov-Bohm
  photonic cages in waveguide and coupled resonator lattices by synthetic
  magnetic fields}},}\ }\href {\doibase 10.1364/OL.39.005892} {\bibfield
  {journal} {\bibinfo  {journal} {Opt. Lett.}\ }\textbf {\bibinfo {volume}
  {39}},\ \bibinfo {pages} {5892--5895} (\bibinfo {year} {2014})}\BibitemShut
  {NoStop}%
\bibitem [{\citenamefont {Kibis}\ \emph {et~al.}(2015)\citenamefont {Kibis},
  \citenamefont {Sigurdsson},\ and\ \citenamefont
  {Shelykh}}]{kibis2015aharonov}%
  \BibitemOpen
  \bibfield  {author} {\bibinfo {author} {\bibfnamefont {O.~V.}\ \bibnamefont
  {Kibis}}, \bibinfo {author} {\bibfnamefont {H.}~\bibnamefont {Sigurdsson}}, \
  and\ \bibinfo {author} {\bibfnamefont {I.~A.}\ \bibnamefont {Shelykh}},\
  }\bibfield  {title} {\enquote {\bibinfo {title} {{Aharonov-Bohm effect for
  excitons in a semiconductor quantum ring dressed by circularly polarized
  light}},}\ }\href {\doibase 10.1103/PhysRevB.91.235308} {\bibfield  {journal}
  {\bibinfo  {journal} {Phys. Rev. B}\ }\textbf {\bibinfo {volume} {91}},\
  \bibinfo {pages} {235308} (\bibinfo {year} {2015})}\BibitemShut {NoStop}%
\bibitem [{\citenamefont {Hasan}\ \emph {et~al.}(2016)\citenamefont {Hasan},
  \citenamefont {Iorsh}, \citenamefont {Kibis},\ and\ \citenamefont
  {Shelykh}}]{hasan2016optically}%
  \BibitemOpen
  \bibfield  {author} {\bibinfo {author} {\bibfnamefont {M.}~\bibnamefont
  {Hasan}}, \bibinfo {author} {\bibfnamefont {I.~V.}\ \bibnamefont {Iorsh}},
  \bibinfo {author} {\bibfnamefont {O.~V.}\ \bibnamefont {Kibis}}, \ and\
  \bibinfo {author} {\bibfnamefont {I.~A.}\ \bibnamefont {Shelykh}},\
  }\bibfield  {title} {\enquote {\bibinfo {title} {Optically controlled
  periodical chain of quantum rings},}\ }\href {\doibase
  10.1103/PhysRevB.93.125401} {\bibfield  {journal} {\bibinfo  {journal} {Phys.
  Rev. B}\ }\textbf {\bibinfo {volume} {93}},\ \bibinfo {pages} {125401}
  (\bibinfo {year} {2016})}\BibitemShut {NoStop}%
\bibitem [{\citenamefont {Fang}\ \emph
  {et~al.}(2012{\natexlab{b}})\citenamefont {Fang}, \citenamefont {Yu},\ and\
  \citenamefont {Fan}}]{fang2012realizing}%
  \BibitemOpen
  \bibfield  {author} {\bibinfo {author} {\bibfnamefont {Kejie}\ \bibnamefont
  {Fang}}, \bibinfo {author} {\bibfnamefont {Zongfu}\ \bibnamefont {Yu}}, \
  and\ \bibinfo {author} {\bibfnamefont {Shanhui}\ \bibnamefont {Fan}},\
  }\bibfield  {title} {\enquote {\bibinfo {title} {Realizing effective magnetic
  field for photons by controlling the phase of dynamic modulation},}\ }\href
  {\doibase 10.1038/nphoton.2012.236} {\bibfield  {journal} {\bibinfo
  {journal} {Nature Photonics}\ }\textbf {\bibinfo {volume} {6}},\ \bibinfo
  {pages} {782--787} (\bibinfo {year} {2012}{\natexlab{b}})}\BibitemShut
  {NoStop}%
\bibitem [{\citenamefont {Mukherjee}\ \emph {et~al.}(2018)\citenamefont
  {Mukherjee}, \citenamefont {Di~Liberto}, \citenamefont {\"Ohberg},
  \citenamefont {Thomson},\ and\ \citenamefont
  {Goldman}}]{mukherjee2018experimental}%
  \BibitemOpen
  \bibfield  {author} {\bibinfo {author} {\bibfnamefont {Sebabrata}\
  \bibnamefont {Mukherjee}}, \bibinfo {author} {\bibfnamefont {Marco}\
  \bibnamefont {Di~Liberto}}, \bibinfo {author} {\bibfnamefont {Patrik}\
  \bibnamefont {\"Ohberg}}, \bibinfo {author} {\bibfnamefont {Robert~R.}\
  \bibnamefont {Thomson}}, \ and\ \bibinfo {author} {\bibfnamefont {Nathan}\
  \bibnamefont {Goldman}},\ }\bibfield  {title} {\enquote {\bibinfo {title}
  {Experimental observation of {Aharonov-Bohm} cages in photonic lattices},}\
  }\href {\doibase 10.1103/PhysRevLett.121.075502} {\bibfield  {journal}
  {\bibinfo  {journal} {Phys. Rev. Lett.}\ }\textbf {\bibinfo {volume} {121}},\
  \bibinfo {pages} {075502} (\bibinfo {year} {2018})}\BibitemShut {NoStop}%
\bibitem [{\citenamefont {Gladchenko}\ \emph {et~al.}(2009)\citenamefont
  {Gladchenko}, \citenamefont {Olaya}, \citenamefont {Dupont-Ferrier},
  \citenamefont {Douçot}, \citenamefont {Ioffe},\ and\ \citenamefont
  {Gershenson}}]{gladchenko2009superconducting}%
  \BibitemOpen
  \bibfield  {author} {\bibinfo {author} {\bibfnamefont {Sergey}\ \bibnamefont
  {Gladchenko}}, \bibinfo {author} {\bibfnamefont {David}\ \bibnamefont
  {Olaya}}, \bibinfo {author} {\bibfnamefont {Eva}\ \bibnamefont
  {Dupont-Ferrier}}, \bibinfo {author} {\bibfnamefont {Benoit}\ \bibnamefont
  {Douçot}}, \bibinfo {author} {\bibfnamefont {Lev~B.}\ \bibnamefont {Ioffe}},
  \ and\ \bibinfo {author} {\bibfnamefont {Michael~E.}\ \bibnamefont
  {Gershenson}},\ }\bibfield  {title} {\enquote {\bibinfo {title}
  {Superconducting nanocircuits for topologically protected qubits},}\ }\href
  {\doibase 10.1038/nphys1151} {\bibfield  {journal} {\bibinfo  {journal}
  {Nature Phys.}\ }\textbf {\bibinfo {volume} {5}},\ \bibinfo {pages} {48--53}
  (\bibinfo {year} {2009})}\BibitemShut {NoStop}%
\bibitem [{\citenamefont {Gligori\ifmmode~\acute{c}\else \'{c}\fi{}}\ \emph
  {et~al.}(2019)\citenamefont {Gligori\ifmmode~\acute{c}\else \'{c}\fi{}},
  \citenamefont {Beli\ifmmode~\check{c}\else \v{c}\fi{}ev}, \citenamefont
  {Leykam},\ and\ \citenamefont {Maluckov}}]{gligoric2018nonlinear}%
  \BibitemOpen
  \bibfield  {author} {\bibinfo {author} {\bibfnamefont {Goran}\ \bibnamefont
  {Gligori\ifmmode~\acute{c}\else \'{c}\fi{}}}, \bibinfo {author}
  {\bibfnamefont {Petra~P.}\ \bibnamefont {Beli\ifmmode~\check{c}\else
  \v{c}\fi{}ev}}, \bibinfo {author} {\bibfnamefont {Daniel}\ \bibnamefont
  {Leykam}}, \ and\ \bibinfo {author} {\bibfnamefont {Aleksandra}\ \bibnamefont
  {Maluckov}},\ }\bibfield  {title} {\enquote {\bibinfo {title} {Nonlinear
  symmetry breaking of {Aharonov-Bohm} cages},}\ }\href {\doibase
  10.1103/PhysRevA.99.013826} {\bibfield  {journal} {\bibinfo  {journal} {Phys.
  Rev. A}\ }\textbf {\bibinfo {volume} {99}},\ \bibinfo {pages} {013826}
  (\bibinfo {year} {2019})}\BibitemShut {NoStop}%
\bibitem [{\citenamefont {Di~Liberto}\ \emph {et~al.}(2019)\citenamefont
  {Di~Liberto}, \citenamefont {Mukherjee},\ and\ \citenamefont
  {Goldman}}]{diliberto2019nonlinear}%
  \BibitemOpen
  \bibfield  {author} {\bibinfo {author} {\bibfnamefont {Marco}\ \bibnamefont
  {Di~Liberto}}, \bibinfo {author} {\bibfnamefont {Sebabrata}\ \bibnamefont
  {Mukherjee}}, \ and\ \bibinfo {author} {\bibfnamefont {Nathan}\ \bibnamefont
  {Goldman}},\ }\bibfield  {title} {\enquote {\bibinfo {title} {Nonlinear
  dynamics of {Aharonov-Bohm} cages},}\ }\href {\doibase
  10.1103/PhysRevA.100.043829} {\bibfield  {journal} {\bibinfo  {journal}
  {Phys. Rev. A}\ }\textbf {\bibinfo {volume} {100}},\ \bibinfo {pages}
  {043829} (\bibinfo {year} {2019})}\BibitemShut {NoStop}%
\bibitem [{\citenamefont {Vidal}\ \emph {et~al.}(2000)\citenamefont {Vidal},
  \citenamefont {Dou\ifmmode~\mbox{\c{c}}\else \c{c}\fi{}ot}, \citenamefont
  {Mosseri},\ and\ \citenamefont {Butaud}}]{vidal2000interaction}%
  \BibitemOpen
  \bibfield  {author} {\bibinfo {author} {\bibfnamefont {Julien}\ \bibnamefont
  {Vidal}}, \bibinfo {author} {\bibfnamefont {Beno\^{\i}t}\ \bibnamefont
  {Dou\ifmmode~\mbox{\c{c}}\else \c{c}\fi{}ot}}, \bibinfo {author}
  {\bibfnamefont {R\'emy}\ \bibnamefont {Mosseri}}, \ and\ \bibinfo {author}
  {\bibfnamefont {Patrick}\ \bibnamefont {Butaud}},\ }\bibfield  {title}
  {\enquote {\bibinfo {title} {Interaction induced delocalization for two
  particles in a periodic potential},}\ }\href {\doibase
  10.1103/PhysRevLett.85.3906} {\bibfield  {journal} {\bibinfo  {journal}
  {Phys. Rev. Lett.}\ }\textbf {\bibinfo {volume} {85}},\ \bibinfo {pages}
  {3906--3909} (\bibinfo {year} {2000})}\BibitemShut {NoStop}%
\bibitem [{\citenamefont {Creutz}(1999)}]{creutz1999end}%
  \BibitemOpen
  \bibfield  {author} {\bibinfo {author} {\bibfnamefont {Michael}\ \bibnamefont
  {Creutz}},\ }\bibfield  {title} {\enquote {\bibinfo {title} {End states,
  ladder compounds, and domain-wall fermions},}\ }\href {\doibase
  10.1103/PhysRevLett.83.2636} {\bibfield  {journal} {\bibinfo  {journal}
  {Phys. Rev. Lett.}\ }\textbf {\bibinfo {volume} {83}},\ \bibinfo {pages}
  {2636--2639} (\bibinfo {year} {1999})}\BibitemShut {NoStop}%
\bibitem [{\citenamefont {Takayoshi}\ \emph {et~al.}(2013)\citenamefont
  {Takayoshi}, \citenamefont {Katsura}, \citenamefont {Watanabe},\ and\
  \citenamefont {Aoki}}]{takayoshi2013phase}%
  \BibitemOpen
  \bibfield  {author} {\bibinfo {author} {\bibfnamefont {Shintaro}\
  \bibnamefont {Takayoshi}}, \bibinfo {author} {\bibfnamefont {Hosho}\
  \bibnamefont {Katsura}}, \bibinfo {author} {\bibfnamefont {Noriaki}\
  \bibnamefont {Watanabe}}, \ and\ \bibinfo {author} {\bibfnamefont {Hideo}\
  \bibnamefont {Aoki}},\ }\bibfield  {title} {\enquote {\bibinfo {title}
  {{Phase diagram and pair Tomonaga-Luttinger liquid in a Bose-Hubbard model
  with flat bands}},}\ }\href {\doibase 10.1103/PhysRevA.88.063613} {\bibfield
  {journal} {\bibinfo  {journal} {Phys. Rev. A}\ }\textbf {\bibinfo {volume}
  {88}},\ \bibinfo {pages} {063613} (\bibinfo {year} {2013})}\BibitemShut
  {NoStop}%
\bibitem [{\citenamefont {Tovmasyan}\ \emph {et~al.}(2013)\citenamefont
  {Tovmasyan}, \citenamefont {van Nieuwenburg},\ and\ \citenamefont
  {Huber}}]{tovmasyan2013geometry}%
  \BibitemOpen
  \bibfield  {author} {\bibinfo {author} {\bibfnamefont {Murad}\ \bibnamefont
  {Tovmasyan}}, \bibinfo {author} {\bibfnamefont {Evert P.~L.}\ \bibnamefont
  {van Nieuwenburg}}, \ and\ \bibinfo {author} {\bibfnamefont {Sebastian~D.}\
  \bibnamefont {Huber}},\ }\bibfield  {title} {\enquote {\bibinfo {title}
  {Geometry-induced pair condensation},}\ }\href {\doibase
  10.1103/PhysRevB.88.220510} {\bibfield  {journal} {\bibinfo  {journal} {Phys.
  Rev. B}\ }\textbf {\bibinfo {volume} {88}},\ \bibinfo {pages} {220510}
  (\bibinfo {year} {2013})}\BibitemShut {NoStop}%
\bibitem [{\citenamefont {Tovmasyan}\ \emph {et~al.}(2016)\citenamefont
  {Tovmasyan}, \citenamefont {Peotta}, \citenamefont {T\"orm\"a},\ and\
  \citenamefont {Huber}}]{tovmasyan2016effective}%
  \BibitemOpen
  \bibfield  {author} {\bibinfo {author} {\bibfnamefont {Murad}\ \bibnamefont
  {Tovmasyan}}, \bibinfo {author} {\bibfnamefont {Sebastiano}\ \bibnamefont
  {Peotta}}, \bibinfo {author} {\bibfnamefont {P\"aivi}\ \bibnamefont
  {T\"orm\"a}}, \ and\ \bibinfo {author} {\bibfnamefont {Sebastian~D.}\
  \bibnamefont {Huber}},\ }\bibfield  {title} {\enquote {\bibinfo {title}
  {Effective theory and emergent $\text{SU}(2)$ symmetry in the flat bands of
  attractive hubbard models},}\ }\href {\doibase 10.1103/PhysRevB.94.245149}
  {\bibfield  {journal} {\bibinfo  {journal} {Phys. Rev. B}\ }\textbf {\bibinfo
  {volume} {94}},\ \bibinfo {pages} {245149} (\bibinfo {year}
  {2016})}\BibitemShut {NoStop}%
\bibitem [{\citenamefont {J\"unemann}\ \emph {et~al.}(2017)\citenamefont
  {J\"unemann}, \citenamefont {Piga}, \citenamefont {Ran}, \citenamefont
  {Lewenstein}, \citenamefont {Rizzi},\ and\ \citenamefont
  {Bermudez}}]{junemann2017exploring}%
  \BibitemOpen
  \bibfield  {author} {\bibinfo {author} {\bibfnamefont {J.}~\bibnamefont
  {J\"unemann}}, \bibinfo {author} {\bibfnamefont {A.}~\bibnamefont {Piga}},
  \bibinfo {author} {\bibfnamefont {S.-J.}\ \bibnamefont {Ran}}, \bibinfo
  {author} {\bibfnamefont {M.}~\bibnamefont {Lewenstein}}, \bibinfo {author}
  {\bibfnamefont {M.}~\bibnamefont {Rizzi}}, \ and\ \bibinfo {author}
  {\bibfnamefont {A.}~\bibnamefont {Bermudez}},\ }\bibfield  {title} {\enquote
  {\bibinfo {title} {Exploring interacting topological insulators with
  ultracold atoms: The synthetic creutz-hubbard model},}\ }\href {\doibase
  10.1103/PhysRevX.7.031057} {\bibfield  {journal} {\bibinfo  {journal} {Phys.
  Rev. X}\ }\textbf {\bibinfo {volume} {7}},\ \bibinfo {pages} {031057}
  (\bibinfo {year} {2017})}\BibitemShut {NoStop}%
\bibitem [{\citenamefont {Tovmasyan}\ \emph {et~al.}(2018)\citenamefont
  {Tovmasyan}, \citenamefont {Peotta}, \citenamefont {Liang}, \citenamefont
  {T\"orm\"a},\ and\ \citenamefont {Huber}}]{tovmasyan2018preformed}%
  \BibitemOpen
  \bibfield  {author} {\bibinfo {author} {\bibfnamefont {Murad}\ \bibnamefont
  {Tovmasyan}}, \bibinfo {author} {\bibfnamefont {Sebastiano}\ \bibnamefont
  {Peotta}}, \bibinfo {author} {\bibfnamefont {Long}\ \bibnamefont {Liang}},
  \bibinfo {author} {\bibfnamefont {P\"aivi}\ \bibnamefont {T\"orm\"a}}, \ and\
  \bibinfo {author} {\bibfnamefont {Sebastian~D.}\ \bibnamefont {Huber}},\
  }\bibfield  {title} {\enquote {\bibinfo {title} {Preformed pairs in flat
  bloch bands},}\ }\href {\doibase 10.1103/PhysRevB.98.134513} {\bibfield
  {journal} {\bibinfo  {journal} {Phys. Rev. B}\ }\textbf {\bibinfo {volume}
  {98}},\ \bibinfo {pages} {134513} (\bibinfo {year} {2018})}\BibitemShut
  {NoStop}%
\bibitem [{\citenamefont {Vidal}\ \emph {et~al.}(2001)\citenamefont {Vidal},
  \citenamefont {Butaud}, \citenamefont {Dou\ifmmode~\mbox{\c{c}}\else
  \c{c}\fi{}ot},\ and\ \citenamefont {Mosseri}}]{vidal2001disorder}%
  \BibitemOpen
  \bibfield  {author} {\bibinfo {author} {\bibfnamefont {Julien}\ \bibnamefont
  {Vidal}}, \bibinfo {author} {\bibfnamefont {Patrick}\ \bibnamefont {Butaud}},
  \bibinfo {author} {\bibfnamefont {Benoit}\ \bibnamefont
  {Dou\ifmmode~\mbox{\c{c}}\else \c{c}\fi{}ot}}, \ and\ \bibinfo {author}
  {\bibfnamefont {R\'emy}\ \bibnamefont {Mosseri}},\ }\bibfield  {title}
  {\enquote {\bibinfo {title} {Disorder and interactions in {Aharonov-Bohm}
  cages},}\ }\href {\doibase 10.1103/PhysRevB.64.155306} {\bibfield  {journal}
  {\bibinfo  {journal} {Phys. Rev. B}\ }\textbf {\bibinfo {volume} {64}},\
  \bibinfo {pages} {155306} (\bibinfo {year} {2001})}\BibitemShut {NoStop}%
\bibitem [{\citenamefont {Danieli}\ \emph {et~al.}(2020)\citenamefont
  {Danieli}, \citenamefont {Andreanov}, \citenamefont {Mithun},\ and\
  \citenamefont {Flach}}]{danieli2020quantum}%
  \BibitemOpen
  \bibfield  {author} {\bibinfo {author} {\bibfnamefont {Carlo}\ \bibnamefont
  {Danieli}}, \bibinfo {author} {\bibfnamefont {Alexei}\ \bibnamefont
  {Andreanov}}, \bibinfo {author} {\bibfnamefont {Thudiyangal}\ \bibnamefont
  {Mithun}}, \ and\ \bibinfo {author} {\bibfnamefont {Sergej}\ \bibnamefont
  {Flach}},\ }\href@noop {} {\enquote {\bibinfo {title} {Quantum caging in
  interacting many-body all-bands-flat lattices},}\ } (\bibinfo {year}
  {2020}),\ \Eprint {http://arxiv.org/abs/2004.11880} {arXiv:2004.11880
  [cond-mat.quant-gas]} \BibitemShut {NoStop}%
\bibitem [{\citenamefont {Read}(2017)}]{read2017compactly}%
  \BibitemOpen
  \bibfield  {author} {\bibinfo {author} {\bibfnamefont {N.}~\bibnamefont
  {Read}},\ }\bibfield  {title} {\enquote {\bibinfo {title} {Compactly
  supported wannier functions and algebraic $k$-theory},}\ }\href {\doibase
  10.1103/PhysRevB.95.115309} {\bibfield  {journal} {\bibinfo  {journal} {Phys.
  Rev. B}\ }\textbf {\bibinfo {volume} {95}},\ \bibinfo {pages} {115309}
  (\bibinfo {year} {2017})}\BibitemShut {NoStop}%
\bibitem [{\citenamefont {Flach}\ \emph {et~al.}(2014)\citenamefont {Flach},
  \citenamefont {Leykam}, \citenamefont {Bodyfelt}, \citenamefont {Matthies},\
  and\ \citenamefont {Desyatnikov}}]{flach2014detangling}%
  \BibitemOpen
  \bibfield  {author} {\bibinfo {author} {\bibfnamefont {Sergej}\ \bibnamefont
  {Flach}}, \bibinfo {author} {\bibfnamefont {Daniel}\ \bibnamefont {Leykam}},
  \bibinfo {author} {\bibfnamefont {Joshua~D.}\ \bibnamefont {Bodyfelt}},
  \bibinfo {author} {\bibfnamefont {Peter}\ \bibnamefont {Matthies}}, \ and\
  \bibinfo {author} {\bibfnamefont {Anton~S.}\ \bibnamefont {Desyatnikov}},\
  }\bibfield  {title} {\enquote {\bibinfo {title} {Detangling flat bands into
  {Fano} lattices},}\ }\href {http://stacks.iop.org/0295-5075/105/i=3/a=30001}
  {\bibfield  {journal} {\bibinfo  {journal} {Europhys. Lett.}\ }\textbf
  {\bibinfo {volume} {105}},\ \bibinfo {pages} {30001} (\bibinfo {year}
  {2014})}\BibitemShut {NoStop}%
\bibitem [{\citenamefont {Baboux}\ \emph {et~al.}(2016)\citenamefont {Baboux},
  \citenamefont {Ge}, \citenamefont {Jacqmin}, \citenamefont {Biondi},
  \citenamefont {Galopin}, \citenamefont {Lema\^{\i}tre}, \citenamefont
  {Le~Gratiet}, \citenamefont {Sagnes}, \citenamefont {Schmidt}, \citenamefont
  {T\"ureci}, \citenamefont {Amo},\ and\ \citenamefont
  {Bloch}}]{baboux2016bosonic}%
  \BibitemOpen
  \bibfield  {author} {\bibinfo {author} {\bibfnamefont {F.}~\bibnamefont
  {Baboux}}, \bibinfo {author} {\bibfnamefont {L.}~\bibnamefont {Ge}}, \bibinfo
  {author} {\bibfnamefont {T.}~\bibnamefont {Jacqmin}}, \bibinfo {author}
  {\bibfnamefont {M.}~\bibnamefont {Biondi}}, \bibinfo {author} {\bibfnamefont
  {E.}~\bibnamefont {Galopin}}, \bibinfo {author} {\bibfnamefont
  {A.}~\bibnamefont {Lema\^{\i}tre}}, \bibinfo {author} {\bibfnamefont
  {L.}~\bibnamefont {Le~Gratiet}}, \bibinfo {author} {\bibfnamefont
  {I.}~\bibnamefont {Sagnes}}, \bibinfo {author} {\bibfnamefont
  {S.}~\bibnamefont {Schmidt}}, \bibinfo {author} {\bibfnamefont {H.~E.}\
  \bibnamefont {T\"ureci}}, \bibinfo {author} {\bibfnamefont {A.}~\bibnamefont
  {Amo}}, \ and\ \bibinfo {author} {\bibfnamefont {J.}~\bibnamefont {Bloch}},\
  }\bibfield  {title} {\enquote {\bibinfo {title} {Bosonic condensation and
  disorder-induced localization in a flat band},}\ }\href {\doibase
  10.1103/PhysRevLett.116.066402} {\bibfield  {journal} {\bibinfo  {journal}
  {Phys. Rev. Lett.}\ }\textbf {\bibinfo {volume} {116}},\ \bibinfo {pages}
  {066402} (\bibinfo {year} {2016})}\BibitemShut {NoStop}%
\bibitem [{\citenamefont {Real}\ \emph {et~al.}(2017)\citenamefont {Real},
  \citenamefont {Cantillano}, \citenamefont {L{\'o}pez-Gonz{\'a}lez},
  \citenamefont {Szameit}, \citenamefont {Aono}, \citenamefont {Naruse},
  \citenamefont {Kim}, \citenamefont {Wang},\ and\ \citenamefont
  {Vicencio}}]{real2017flat}%
  \BibitemOpen
  \bibfield  {author} {\bibinfo {author} {\bibfnamefont {Basti{\'a}n}\
  \bibnamefont {Real}}, \bibinfo {author} {\bibfnamefont {Camilo}\ \bibnamefont
  {Cantillano}}, \bibinfo {author} {\bibfnamefont {Dany}\ \bibnamefont
  {L{\'o}pez-Gonz{\'a}lez}}, \bibinfo {author} {\bibfnamefont {Alexander}\
  \bibnamefont {Szameit}}, \bibinfo {author} {\bibfnamefont {Masashi}\
  \bibnamefont {Aono}}, \bibinfo {author} {\bibfnamefont {Makoto}\ \bibnamefont
  {Naruse}}, \bibinfo {author} {\bibfnamefont {Song-Ju}\ \bibnamefont {Kim}},
  \bibinfo {author} {\bibfnamefont {Kai}\ \bibnamefont {Wang}}, \ and\ \bibinfo
  {author} {\bibfnamefont {Rodrigo~A}\ \bibnamefont {Vicencio}},\ }\bibfield
  {title} {\enquote {\bibinfo {title} {Flat-band light dynamics in stub
  photonic lattices},}\ }\href
  {https://www.nature.com/articles/s41598-017-15441-2} {\bibfield  {journal}
  {\bibinfo  {journal} {Scientific Reports}\ }\textbf {\bibinfo {volume} {7}},\
  \bibinfo {pages} {15085} (\bibinfo {year} {2017})}\BibitemShut {NoStop}%
\bibitem [{\citenamefont {Bergman}\ \emph {et~al.}(2008)\citenamefont
  {Bergman}, \citenamefont {Wu},\ and\ \citenamefont
  {Balents}}]{bergman2008band}%
  \BibitemOpen
  \bibfield  {author} {\bibinfo {author} {\bibfnamefont {Doron~L.}\
  \bibnamefont {Bergman}}, \bibinfo {author} {\bibfnamefont {Congjun}\
  \bibnamefont {Wu}}, \ and\ \bibinfo {author} {\bibfnamefont {Leon}\
  \bibnamefont {Balents}},\ }\bibfield  {title} {\enquote {\bibinfo {title}
  {Band touching from real-space topology in frustrated hopping models},}\
  }\href {\doibase 10.1103/PhysRevB.78.125104} {\bibfield  {journal} {\bibinfo
  {journal} {Phys. Rev. B}\ }\textbf {\bibinfo {volume} {78}},\ \bibinfo
  {pages} {125104} (\bibinfo {year} {2008})}\BibitemShut {NoStop}%
\bibitem [{\citenamefont {Dias}\ and\ \citenamefont
  {Gouveia}(2015)}]{dias2015origami}%
  \BibitemOpen
  \bibfield  {author} {\bibinfo {author} {\bibfnamefont {R.~G.}\ \bibnamefont
  {Dias}}\ and\ \bibinfo {author} {\bibfnamefont {J.~D.}\ \bibnamefont
  {Gouveia}},\ }\bibfield  {title} {\enquote {\bibinfo {title} {Origami rules
  for the construction of localized eigenstates of the {Hubbard} model in
  decorated lattices},}\ }\href {http://dx.doi.org/10.1038/srep16852}
  {\bibfield  {journal} {\bibinfo  {journal} {Sci. Rep.}\ }\textbf {\bibinfo
  {volume} {5}},\ \bibinfo {pages} {16852 EP --} (\bibinfo {year}
  {2015})}\BibitemShut {NoStop}%
\bibitem [{\citenamefont {Morales-Inostroza}\ and\ \citenamefont
  {Vicencio}(2016)}]{morales2016simple}%
  \BibitemOpen
  \bibfield  {author} {\bibinfo {author} {\bibfnamefont {Luis}\ \bibnamefont
  {Morales-Inostroza}}\ and\ \bibinfo {author} {\bibfnamefont {Rodrigo~A.}\
  \bibnamefont {Vicencio}},\ }\bibfield  {title} {\enquote {\bibinfo {title}
  {Simple method to construct flat-band lattices},}\ }\href {\doibase
  10.1103/PhysRevA.94.043831} {\bibfield  {journal} {\bibinfo  {journal} {Phys.
  Rev. A}\ }\textbf {\bibinfo {volume} {94}},\ \bibinfo {pages} {043831}
  (\bibinfo {year} {2016})}\BibitemShut {NoStop}%
\bibitem [{\citenamefont {Maimaiti}\ \emph {et~al.}(2017)\citenamefont
  {Maimaiti}, \citenamefont {Andreanov}, \citenamefont {Park}, \citenamefont
  {Gendelman},\ and\ \citenamefont {Flach}}]{maimaiti2017compact}%
  \BibitemOpen
  \bibfield  {author} {\bibinfo {author} {\bibfnamefont {Wulayimu}\
  \bibnamefont {Maimaiti}}, \bibinfo {author} {\bibfnamefont {Alexei}\
  \bibnamefont {Andreanov}}, \bibinfo {author} {\bibfnamefont {Hee~Chul}\
  \bibnamefont {Park}}, \bibinfo {author} {\bibfnamefont {Oleg}\ \bibnamefont
  {Gendelman}}, \ and\ \bibinfo {author} {\bibfnamefont {Sergej}\ \bibnamefont
  {Flach}},\ }\bibfield  {title} {\enquote {\bibinfo {title} {Compact localized
  states and flat-band generators in one dimension},}\ }\href {\doibase
  10.1103/PhysRevB.95.115135} {\bibfield  {journal} {\bibinfo  {journal} {Phys.
  Rev. B}\ }\textbf {\bibinfo {volume} {95}},\ \bibinfo {pages} {115135}
  (\bibinfo {year} {2017})}\BibitemShut {NoStop}%
\bibitem [{\citenamefont {R\"ontgen}\ \emph {et~al.}(2018)\citenamefont
  {R\"ontgen}, \citenamefont {Morfonios},\ and\ \citenamefont
  {Schmelcher}}]{rontgen2018compact}%
  \BibitemOpen
  \bibfield  {author} {\bibinfo {author} {\bibfnamefont {M.}~\bibnamefont
  {R\"ontgen}}, \bibinfo {author} {\bibfnamefont {C.~V.}\ \bibnamefont
  {Morfonios}}, \ and\ \bibinfo {author} {\bibfnamefont {P.}~\bibnamefont
  {Schmelcher}},\ }\bibfield  {title} {\enquote {\bibinfo {title} {Compact
  localized states and flat bands from local symmetry partitioning},}\ }\href
  {\doibase 10.1103/PhysRevB.97.035161} {\bibfield  {journal} {\bibinfo
  {journal} {Phys. Rev. B}\ }\textbf {\bibinfo {volume} {97}},\ \bibinfo
  {pages} {035161} (\bibinfo {year} {2018})}\BibitemShut {NoStop}%
\bibitem [{\citenamefont {Toikka}\ and\ \citenamefont
  {Andreanov}(2018)}]{toikka2018necessary}%
  \BibitemOpen
  \bibfield  {author} {\bibinfo {author} {\bibfnamefont {L~A}\ \bibnamefont
  {Toikka}}\ and\ \bibinfo {author} {\bibfnamefont {A}~\bibnamefont
  {Andreanov}},\ }\bibfield  {title} {\enquote {\bibinfo {title} {Necessary and
  sufficient conditions for flat bands in m-dimensional n-band lattices with
  complex-valued nearest-neighbour hopping},}\ }\href {\doibase
  10.1088/1751-8121/aaf25c} {\bibfield  {journal} {\bibinfo  {journal} {J Phys.
  A: Math. Theor}\ }\textbf {\bibinfo {volume} {52}},\ \bibinfo {pages}
  {02LT04} (\bibinfo {year} {2018})}\BibitemShut {NoStop}%
\bibitem [{\citenamefont {Maimaiti}\ \emph {et~al.}(2019)\citenamefont
  {Maimaiti}, \citenamefont {Flach},\ and\ \citenamefont
  {Andreanov}}]{maimaiti2019universal}%
  \BibitemOpen
  \bibfield  {author} {\bibinfo {author} {\bibfnamefont {Wulayimu}\
  \bibnamefont {Maimaiti}}, \bibinfo {author} {\bibfnamefont {Sergej}\
  \bibnamefont {Flach}}, \ and\ \bibinfo {author} {\bibfnamefont {Alexei}\
  \bibnamefont {Andreanov}},\ }\bibfield  {title} {\enquote {\bibinfo {title}
  {Universal $d=1$ flat band generator from compact localized states},}\ }\href
  {\doibase 10.1103/PhysRevB.99.125129} {\bibfield  {journal} {\bibinfo
  {journal} {Phys. Rev. B}\ }\textbf {\bibinfo {volume} {99}},\ \bibinfo
  {pages} {125129} (\bibinfo {year} {2019})}\BibitemShut {NoStop}%
\bibitem [{\citenamefont {Haldane}(1988)}]{haldane1988model}%
  \BibitemOpen
  \bibfield  {author} {\bibinfo {author} {\bibfnamefont {F.~D.~M.}\
  \bibnamefont {Haldane}},\ }\bibfield  {title} {\enquote {\bibinfo {title}
  {Model for a quantum hall effect without landau levels: Condensed-matter
  realization of the "parity anomaly"},}\ }\href {\doibase
  10.1103/PhysRevLett.61.2015} {\bibfield  {journal} {\bibinfo  {journal}
  {Phys. Rev. Lett.}\ }\textbf {\bibinfo {volume} {61}},\ \bibinfo {pages}
  {2015--2018} (\bibinfo {year} {1988})}\BibitemShut {NoStop}%
\bibitem [{Note1()}]{Note1}%
  \BibitemOpen
  \bibinfo {note} {Consider a gauge transformation $\protect \mathcal {G}:
  \protect \{a_n,b_n\protect \} \DOTSB \mapstochar \DOTSB \protect \relbar
  \protect \joinrel \rightarrow \protect \{e^{i\xi _n} a_n, e^{i\eta _n}
  b_n\protect \}$ defined for arbitrary phase functions $\xi _n,\eta _n$. The
  submanifold of lattices related to the Creutz ladder via $\protect \mathcal
  {G}$ is given by the family of parameters $z_1 = e^{i\protect \frac {\xi
  _n-\eta _n}{2}}/\protect \sqrt {2}, \ w_1 = i e^{i\protect \frac {\xi _n-\eta
  _n}{2}} /\protect \sqrt {2}, \ z_2 = e^{i g_n(\xi _n , \eta _n)}/\protect
  \sqrt {2},\ w_2 = e^{-i g_n(\xi _n, \eta _n)}/\protect \sqrt {2}$ in
  Eqs.~(\ref {eq:H0H1_rot0}-\ref {eq:H0H1_rot1}) for arbitrary functions
  $g_n$.}\BibitemShut {Stop}%
\bibitem [{Note2()}]{Note2}%
  \BibitemOpen
  \bibinfo {note} {One could potentially imagine further fine-tuning of the
  interaction that might cancel the total sum but not the individual terms:
  this is however only possible for a specific combination of the amplitudes
  $\phi _{n,a}$, implying that at best only specific excitations can be caged.
  We are interested in caging of any initial excitation and therefore do not
  investigate further this possibility.}\BibitemShut {Stop}%
\bibitem [{Note3()}]{Note3}%
  \BibitemOpen
  \bibinfo {note} {We are relying here on the requirement that any initial
  excitations are caged, so that we do not need to worry about some
  particularly small excitations, where this argument might fail.}\BibitemShut
  {Stop}%
\bibitem [{\citenamefont {Danieli}\ \emph {et~al.}(2019)\citenamefont
  {Danieli}, \citenamefont {Manda}, \citenamefont {Mithun},\ and\ \citenamefont
  {Skokos}}]{danieli2019computational}%
  \BibitemOpen
  \bibfield  {author} {\bibinfo {author} {\bibfnamefont {C.}~\bibnamefont
  {Danieli}}, \bibinfo {author} {\bibfnamefont {B.~Many}\ \bibnamefont
  {Manda}}, \bibinfo {author} {\bibfnamefont {T.}~\bibnamefont {Mithun}}, \
  and\ \bibinfo {author} {\bibfnamefont {Ch.}\ \bibnamefont {Skokos}},\
  }\bibfield  {title} {\enquote {\bibinfo {title} {Computational efficiency of
  numerical integration methods for the tangent dynamics of many-body
  hamiltonian systems in one and two spatial dimensions},}\ }\href {\doibase
  http://dx.doi.org/10.3934/mine.2019.3.447} {\bibfield  {journal} {\bibinfo
  {journal} {Mathematics in Engineering}\ }\textbf {\bibinfo {volume} {1}},\
  \bibinfo {pages} {447} (\bibinfo {year} {2019})}\BibitemShut {NoStop}%
\bibitem [{\citenamefont {R\"ontgen}\ \emph {et~al.}(2019)\citenamefont
  {R\"ontgen}, \citenamefont {Morfonios}, \citenamefont {Brouzos},
  \citenamefont {Diakonos},\ and\ \citenamefont
  {Schmelcher}}]{rontgen2019quantum}%
  \BibitemOpen
  \bibfield  {author} {\bibinfo {author} {\bibfnamefont {M.}~\bibnamefont
  {R\"ontgen}}, \bibinfo {author} {\bibfnamefont {C.~V.}\ \bibnamefont
  {Morfonios}}, \bibinfo {author} {\bibfnamefont {I.}~\bibnamefont {Brouzos}},
  \bibinfo {author} {\bibfnamefont {F.~K.}\ \bibnamefont {Diakonos}}, \ and\
  \bibinfo {author} {\bibfnamefont {P.}~\bibnamefont {Schmelcher}},\ }\bibfield
   {title} {\enquote {\bibinfo {title} {Quantum network transfer and storage
  with compact localized states induced by local symmetries},}\ }\href
  {\doibase 10.1103/PhysRevLett.123.080504} {\bibfield  {journal} {\bibinfo
  {journal} {Phys. Rev. Lett.}\ }\textbf {\bibinfo {volume} {123}},\ \bibinfo
  {pages} {080504} (\bibinfo {year} {2019})}\BibitemShut {NoStop}%
\end{thebibliography}%

\end{document}